\documentclass[reqno,11pt]{amsart}
\usepackage[linktocpage]{hyperref}
\usepackage{graphicx}
\usepackage{slashed}
\usepackage{amscd}
\usepackage{tensor}
\usepackage{amssymb}
\usepackage{esint}
\usepackage{enumitem}
\usepackage{accents}
\usepackage[usenames,dvipsnames]{pstricks}
\usepackage{pst-grad} % For gradients
\usepackage{pst-plot} % For axes
\usepackage[mathscr]{eucal}
\numberwithin{equation}{section}
\allowdisplaybreaks[1]

\title[Incompatibility of frequency splitting and spatial localization]{Incompatibility of Frequency Splitting \\ and Spatial Localization: \\ A Quantitative Analysis of Hegerfeldt's Theorem}
\author[F.\ Finster]{Felix Finster}%May 2020 / January 2022}
\address{Fakult\"at f\"ur Mathematik \\ Universit\"at Regensburg \\ D-93040 Regensburg \\ Germany}
\email{finster@ur.de}
\author[C.F.\ Paganini]{Claudio F.\ Paganini \\ \\ May 2020 / June 2022}
\address{Fakult\"at f\"ur Mathematik \\ Universit\"at Regensburg \\ D-93040 Regensburg \\ Germany}
\address{Max Planck Institute for Gravitational Physics (Albert Einstein Institute), Am M\"uh\-len\-berg 1, D-14476 Potsdam, Germany}
\email{claudio.paganini@ur.com}

\newtheorem{Def}{Definition}[section]
\newtheorem{Thm}[Def]{Theorem}

\newtheorem{Prp}[Def]{Proposition}
\newtheorem{Lemma}[Def]{Lemma}

\newtheorem{Corollary}[Def]{Corollary}

\newcommand{\Thanks}{\vspace*{.5em} \noindent \thanks}
\newcommand{\beq}{\begin{equation}}
\newcommand{\eeq}{\end{equation}}
\newcommand{\Proof}{\begin{proof}}
\newcommand{\QED}{\end{proof} \noindent}

\newcommand{\la}{\langle}
\newcommand{\ra}{\rangle}

\newcommand{\C}{\mathbb{C}}
\newcommand{\R}{\mathbb{R}}

\newcommand{\N}{\mathbb{N}}

\renewcommand{\O}{{\mathscr{O}}}

\DeclareMathOperator{\supp}{supp}

\DeclareMathOperator{\re}{Re}
\DeclareMathOperator{\im}{Im}

\newcommand{\scrM}{\mycal M}

\newcommand{\bitem}{\begin{itemize}[leftmargin=2em]}
\newcommand{\eitem}{\end{itemize}}

\newcommand{\even}{\text{\tiny{\rm{even}}}}
\newcommand{\odd}{\text{\tiny{\rm{odd}}}}
\newcommand{\eo}{\bullet}

\DeclareFontFamily{OT1}{rsfso}{}
\DeclareFontShape{OT1}{rsfso}{m}{n}{ <-7> rsfso5 <7-10> rsfso7 <10-> rsfso10}{}
\DeclareMathAlphabet{\mycal}{OT1}{rsfso}{m}{n}

\begin{document}

\maketitle

\begin{abstract}
We prove quantitative versions of the following statement: If a solution of the $1+1$-dimensional
wave equation has spatially compact support and consists mainly of positive frequencies,
then it must have a significant high-frequency component.
Similar results are proven for the $3+1$-dimensional wave equation.
\end{abstract}

\tableofcontents

\section{Introduction}
The present paper provides a quantitative analysis of a problem that has been studied by different communities in different contexts. On the one hand, in {\em{quantum theory}} it is well-known that spatial localization is
incompatible with the Hamiltonian (i.e., the generator of time translations)  to be bounded from below.
This result, often referred to as {\em{Hegerfeldt's theorem}},
means physically that a quantum system either propagates with infinite
speed (thus violating causality), or else it must involve pair creation or annihilation processes
as described by wave functions involving arbitrarily large negative frequencies\footnote{In order to avoid confusion for readers with a more mathematical background,
we note that, here and throughout the paper, by {\em{frequency}} we always
refer to oscillations in the time variable (in contrast, wave vectors in the spatial variables
are referred to as {\em{momenta}}).}.
Hegerfeldt's theorem has far-reaching consequences for our understanding of the interplay
between locality and the distribution of energy in spacetime.
To give a simple example, it explains why the Feynman propagator~$G_\text{F}(x,y)$
(defined by the condition that ``positive frequencies travel to the future'' and ``negative frequencies travel
to the past'') cannot be causal but instead must have non-vanishing contributions for a large spacelike
separation of~$x$ and~$y$. 

From the point of view of {\em{harmonic analysis}}, on the other hand, Hegerfeldt's theorem can be regarded
as an application of a classic theorem by F.\ and M.\ Riesz, a discussion of which can be found for example in~\cite[Section~I.1]{havin2012uncertainty}.
It constitutes a special case of an {\em{annihilating pair of sets for the Fourier transform}} as discussed in \cite[Section 1.2.1]{havin2001uncertainty}. For related problems in harmonic analysis see for example  \cite{muscalu2013classical} but also \cite{tao2010epsilon}, which contains a power series argument similar to the one we develop in the course of our work in Section \ref{sechighcoeff}.

The proof of Hegerfeldt's theorem (see~\cite{hegerfeldt1974remark} or the concise review
in~\cite[Theorem~3 in Section~4]{castrigiano})
uses complex continuation and the Schwarz reflection principle. This method is general and elegant,
but unfortunately it does not give quantitative information on the frequency splitting.
The goal of the present paper is to prove quantitative versions of Hegerfeldt's theorem.
In order to make the paper accessible to a broader readership, we formulate the problem
and our results purely in the language of hyperbolic 
partial differential equations (PDEs).
From this perspective, Hegerfeldt's theorem states that solutions of hyperbolic PDEs
which have spatially compact support cannot be composed purely of positive (or similarly negative)
frequencies (a clear and detailed proof in the PDE language is given in~\cite[Section~1.8]{thaller}
or~\cite[Corollary 3.6]{CB}).
The quantification we have in mind is the following:
Suppose that at an initial time, a solution has compact support in a ball of radius~$r$.
What can one infer on the possible frequency distributions of the solution?
In particular, how small can the component of negative (or similarly positive) frequency be?

Before making this question mathematically precise and stating our results, we
give an overview of the literature on localization in quantum theory.
The problem of localization in quantum theory has a long history
(see for example \cite{wightman1955configuration} for an overview of the early literature). It was on that backdrop that Hegerfeldt~\cite{hegerfeldt1974remark} proved in 1974 that a quantum mechanical system cannot be localized, or, if initially localized, will spread instantly and thus violate strong Einstein causality.
Skagerstam~\cite{skagerstam1976some} proved the same result with a different method. In particular, he provides an independent proof in the Heisenberg picture. A different attempt at localization using current density four-vectors was pursued in~\cite{gerlach1967kritische,gerlach1967konstruktion}. Hegerfeldt's results were generalized by several authors \cite{perez1977localization, hegerfeldt1980remarks,hegerfeldt1985violation}. In a series of later articles~\cite{hegerfeldt1998instantaneous,hegerfeldt1998causality,hegerfeldt2001particle}, Hegerfeldt discussed these results and their observational consequences in greater detail. Hegerfeldt's theorem has applications to quantum theory in the context of causal
localizations (see for example~\cite{leiseifer, castrigiano} and the references therein for more recent developments).
In~\cite{hegerfeldt2001particle} Hegerfeldt addresses the question why the Dirac equation is not a counter example: the original result is based on the assumption that the Hamiltonian of the system is positive definite,
which obviously is not the case for the Dirac Hamiltonian.
The fact that localized solutions to the Dirac equation always contain contributions of positive and negative
energy has been linked~\cite{hegerfeldt1998instantaneous} to the insight from the field-theoretic perspective that an effective particle corresponds to a ``dressed'' state, i.e.\ that it is surrounded by a cloud of ``virtual'' particle-antiparticle pairs. The appearance of contributions of both positive and negative frequencies in a localized solution to the Dirac equation can be thought of as the PDE counterpart to this phenomenon.    

In the PDE literature, questions similar to those considered in the context of localization in quantum theory were addressed in~\cite{masuda1967unique, masuda1968unique,reed+simon4} in terms of {\em{unique continuation theorems}}, i.e.\ statements of the type that if a solution to a PDE of interest (namely the Schrödinger equation in~\cite{masuda1967unique} or the scalar wave equation in~\cite{masuda1968unique}) vanishes in an open region, then it vanishes everywhere, provided that one requires the solution to be in a suitable regularity class.
%Our results can in fact be understood as quantified unique continuation theorems involving the Fourier transform
%(the connection will be explained further at the end of Section~\ref{secexample}).
Furthermore, see~\cite{baerlocalize, baer+strohmaier} for related results on a Riemannian manifold and~\cite[Section~13]{reed+simon4} for a discussion of similar results for the Schrödinger equation with a potential. It should be noted that, although these results are clearly related, the formulation of the PDE problem does not immediately translate to 
the formulation of the problem of localization in quantum mechanics. The PDE problem assumes the vanishing of a function in a certain domain, while the problem of localization in quantum mechanics assumes that the expectation value of a self-adjoint operator, which is associated to a certain spatial region, vanishes. 

We now specify the mathematical problem and state our main results.
For simplicity, we restrict attention throughout to the cases of 
the scalar wave equation in one and three spatial dimensions.
But, as will become clear from our analysis, our methods also apply 
to other dimensions as well as to the Klein-Gordon equation.
Moreover, our results immediately apply to the equations of higher spin (Maxwell, Dirac, Rarita-Schwinger,
linearized gravity), simply because  in Minkowski space, each component of 
a solution to these equations satisfies the scalar wave equation or Klein-Gordon equation.

In preparation, let us consider the following question:
\bitem
\item[(A)]  \label{questionA}
Assume that at some time~$t_0$, a wave~$\phi(t,x)$ is spatially supported inside a ball of
radius~$r$. Does this imply an a-priori bound for the ratio
\beq \label{quot}
\frac{E(\phi_+)}{E(\phi_-)} 
\eeq
of the energies of the components of positive and negative frequency?
(For notational details see Section~\ref{secprelim}.)
\eitem
The answer to this question is {\em{no}}. Indeed, by making the absolute value of the frequencies of~$\phi$
sufficiently large, one can make the quotient~\eqref{quot} arbitrarily large or small
(for more details see Section~\ref{secexample}).
But, turning this argument around, one concludes that if the quotient~\eqref{quot}
is small, then the wave should have significant high-frequency contributions.
The goal of this paper is to quantify this statement by results of the following form:

\begin{Thm} \label{thmhaupt}
Let~$\phi(t,x)$ be a solution of the scalar wave equation which at some time~$t_0$ is
supported inside a ball of radius~$r>0$,
\[ \supp \phi(t_0, .) \subset B_r(0) \:. \]
Assume that the inequality
\[ E(\phi_-) \leq \varepsilon^2\, E(\phi) \]
holds for some~$\varepsilon \in (0,1]$. Then there is an a-priori estimate for the momentum
distribution of~$\phi$ of the form
\beq \label{main}
\big|k\,\hat{\phi}(k) \big| + \big| \partial_t \hat{\phi}(k) \big| \leq R\big(\varepsilon, r \,|k| \big) \,\sqrt{r\,E(\phi)}\:.
\eeq
\end{Thm} \noindent
Here~$\hat{\phi}$ denotes the spatial Fourier transform (for details see again Section~\ref{secprelim}).

The dispersion relation for the wave equation yields that
frequency and momentum coincide up to a sign.
Therefore, the inequality~\eqref{main} also tells us about the frequency distribution.
By direct computation or using a dimensional argument, one readily
verifies that the inequality~\eqref{main} is scaling invariant. With this in mind, we can always restrict
attention to the case~$r=1$ of a unit ball.
We shall derive several closed expressions for the function~$R$
(see Theorem~\ref{maintheoremsimple}, Theorem~\ref{thmmain2} and 
Corollary~\ref{cormain}, where we always set~$\omega=|k|$).
All these expressions vanish in the limit~$\varepsilon \searrow 0$,
\[ \lim_{\varepsilon \searrow 0} R\big(\varepsilon, |k| \big) = 0 \qquad \text{for all~$k$}\:, \]
as needed for the correspondence to Hegerfeldt's theorem.
If~$\varepsilon$ is positive and small, the inequality~\eqref{main} implies that~$\hat{\phi}(k)$
is small unless~$|k|$ is large. This can be understood as a form of unique continuation, in the sense that, assuming the Fourier transform to have relatively small $L^2$~mass for negative frequencies, we show that the absolute value of the Fourier transform has to be small for small positive frequencies. For partial differential equations,
unique continuation theorems of a similar spirit can be found in~\cite{tataru2004unique,logunov2019lecture}.
There are also related unique continuation results for the Hilbert transform as given for example in~\cite{ruland2019quantitative, alaifari2016lower}.
However, in contrast to these results, it is a specific feature of our method that
we aim at getting uniform estimates for all values of the two parameters~$\varepsilon$
and~$k$. It is one of our main goals to unravel the functional dependence on
these two parameters.

We begin with simple but rough bounds that give a good first
understanding of the underlying mechanism and might be sufficient for some applications.
In the subsequent, more technical parts of the paper we
show that our estimate of the series expansion of the Fourier transform is a solution of a Goursat-Problem,
and employing stationary phase techniques will give rise to significantly improved upper bounds.

In contrast to Hegerfeldt's approach, our methods do not rely on complex analysis. Instead, 
working with Legendre polynomials, we derive estimates for each Taylor coefficient of the
Fourier transform.
From that we infer explicit upper bounds for the Fourier transform at low frequencies. Hegerfeldt's result is obtained in the present considerations by the fact that if we take the limiting case when the compactly supported solution is supported only in the positive frequencies, then the Fourier transform vanishes everywhere, and thus the
function itself is trivial.

We finally note that we expect that our methods and results apply in a much more general setting.
One possible extension is to higher dimensions, as we here illustrate by deriving estimates for
every angular momentum mode of the wave equation in three spatial dimensions.
Moreover, the assumption of compact support could probably be replaced by suitable decay
assumptions of the initial data. Finally, our results should apply to massive equations, to situations in the
presence of external potentials and to equations in curved spacetimes.
Another possible extension would be to consider different decompositions of momentum space into
two subsets which generalize the notions of positive and negative frequencies.
However, these extensions and generalizations go beyond the scope of the present paper.

The paper is structured as follows. In Section~\ref{secprelim} we introduce the mathematical setup and
fix our notation. In Section~\ref{secexample} we discuss a simple example.
The main part of the paper is concerned with the one-dimensional wave equation
(Section~\ref{sec11dim}). 
After recalling a simple pointwise estimate of the Fourier transform (Section~\ref{secpointwise}),
we expand the Fourier transform in a power series (Section~\ref{sectaylormom})
and derive simple estimates of the Taylor coefficients in terms of the energy (Section~\ref{secsimpes}).
In order to derive refined estimates, we decompose the Fourier series into a polynomial
and the remainder. The coefficients of the polynomial are bounded
using $L^2$-estimates together with properties of Legendre polynomials (Section~\ref{sechighcoeff}),
whereas the remainder can be treated with the simple estimates (Section~\ref{secsmalltaylor}).
This gives improved estimates of all Taylor coefficients (see Proposition~\ref{prpanes})
which give rise to estimate the energy distribution of the initial data
in terms of a series~$g(\varepsilon, \omega)$ (see Proposition~\ref{prop:series} in Section~\ref{secsmallinit}).
We proceed with a few simple estimates of this series (Sections~\ref{secsimple} and~\ref{sec48}),
which might be sufficient for future applications and are addressed more towards the theoretical physics community.

The key for getting better estimates of this series is the observation that, as a function
of~$\varepsilon$ and~$\omega$, the series can be transformed to a solution of a 
characteristic initial value problem (Goursat problem) for the $1+1$-dimensional
Klein-Gordon equation (Section~\ref{secgoursat}).
After bringing the initial data into a more explicit form (Section~\ref{secesinit}),
we can solve the Goursat problem with the help of the Klein-Gordon Green's operator
and its representation in momentum space to obtain a contour integral (Section~\ref{seccontour}).
This contour integral can be estimated with a saddle point approximation
and rigorous error bounds~(Section~\ref{secescont}).
It remains to integrate over two parameters which came up in our constructions:
the spatial momentum $k$ (Section~\ref{seckintegral}) and the parameter~$s$
used for the construction of the initial data (Section~\ref{secsintegral}). We thus obtain the improved estimate
for~$g(\omega)$ in Theorem~\ref{prpfinal}. This section contains a number of interesting technical
results and is addressed more at the mathematical community.
Finally, in Section \ref{sec31dim} we extend the results to each angular mode
of the $(3+1)$-dimensional wave equation (see Theorem~\ref{thmmain3}).
The appendix provides an alternative derivation of an integral representation
of the solutions of the Goursat problem given in Section~\ref{secgoursat}.

\section{Preliminaries} \label{secprelim}
\subsection{Fourier Transform}
We recall a well-known result, which is an immediate consequence of
the Paley-Wiener theorem (see~\cite[Section~VI.4]{yosida} or~\cite[Theorem~IX.11]{reed+simon2}).

\begin{Lemma} \label{lemma1}
Let~$\phi \in C^\infty_0(B_1(0))$ be a smooth real- or complex-valued function
with compact support in the interval~$(-1,1) \subset \R$. Then its 
Fourier transform\footnote{We define the Fourier transform with a factor of one and the inverse Fourier transform with a factor of~$1/(2\pi)$.}
\beq \label{fdef1}
\hat{\phi}(k) = \int_{B_1} \phi(x) \: e^{-i kx}\: dx
\eeq
can be represented as a power series
\beq \label{phiser}
\hat{\phi}(k) = \sum_{n=0}^\infty c_n\: k^n \:,
\eeq
with coefficients~$(c_n)_{n\in\N_0}$ bounded by
\begin{align}
|c_n| &\leq \frac{\sqrt{2}}{n!}\: \|\phi\|_{L^2(B_1)} \label{b1} \\
|c_n| &\leq \frac{\sqrt{2}}{(n+1)!}\: \|\partial_x \phi\|_{L^2(B_1)} \label{b2} \:.
\end{align}
\end{Lemma}
\Proof Differentiating~\eqref{fdef1}, we obtain
\[ \big| \hat{\phi}^{(n)}(k) \big| \leq \bigg| \int_{B_1} (-ix)^n \,\phi(x) \: e^{-i kx}\: dx \bigg|
\leq \int_{B_1} \big|\phi(x)\big| \: dx \leq \sqrt{2}\: \|\phi\|_{L^2(B_1)} \:. \]
In particular, setting~$k=0$ we obtain
\[ \big|c_n\big|\: n! = \big| \hat{\phi}^{(n)}(0) \big| \leq \sqrt{2}\: \|\phi\|_{L^2(B_1)} \:, \]
giving the desired bound~\eqref{b1}. Moreover, we conclude that the
Taylor series converges absolutely.

In order to derive~\eqref{b2}, we consider similarly the Fourier transform of the derivative of~$\phi(x)$
to obtain
\[ i k\, \hat{\phi}(k) = \sum_{n=1}^\infty d_n\, k^n \qquad \text{with} \qquad
|d_n| \leq \frac{\sqrt{2}}{n!}\: \|\partial_x\phi\|_{L^2(B_1)} \:. \]
Comparing the last equation with~\eqref{phiser}, one sees that~$c_n=-i d_{n+1}$, giving~\eqref{b2}.
\QED
This estimate shows in particular that~$\hat{\phi}(k)$ is real analytic.

\subsection{Green's Operators and the Causal Fundamental Solution}
The proof of our main theorem is based on estimates of a solution of the
Klein-Gordon equation in $1+1$ dimensions (for details see Section~\ref{secgoursat}).
We now recall the basics on Green's operators needed for this analysis.
The Klein-Gordon equation for a wave~$\phi$ of mass~$m \geq 0$ reads
\[ \big( \partial_t^2 - \partial_x^2 + m^2 \big) \, \phi(t,x) = 0 \:. \]
Green's kernels are distributional solutions of this equation with 
a $\delta$-distribution as inhomogeneity. More precisely, they are defined by the equation
\beq \label{greenpos}
\big( \partial_t^2 - \partial_x^2 + m^2 \big) \, S_{m^2}(t,x) = -\delta(t)\: \delta(x)\:.
\eeq
The Green's operator~$S_{m^2}$ is the corresponding integral operator defined by
\beq \label{greenop}
(S \phi)(t,x) := \int_{\R^2} S_{m^2}(t-t', x-x')\: \phi(t',x')\: dt'\, dx' \:.
\eeq

We now compute the Green's kernel with Fourier methods.
Taking the Fourier transform of the Green's kernel,
\[ %\label{Sfourier}
S_{m^2}(t,x) = \int_{\R^2} \frac{d\omega\: dk}{(2 \pi)^2}\: \hat{S}_{m^2}(\omega, k)\: e^{-i \omega t+ i k x} \:, \]
the differential equation~\eqref{greenpos} reduces to the algebraic equation
\[ (\omega^2 - k^2-m^2)\: \hat{S}(\omega,k) = 1\:. \]
When solving this equation, one must treat the zeros of the function~$\omega^2 - k^2-m^2$
with a suitable deformation in the complex plane. For our purposes, it is useful to choose
\beq \label{Sk}
\begin{split}
\hat{S}_{m^2}^\vee(\omega,k) &= \lim_{\varepsilon \searrow 0} \frac{1}{\omega^2-k^2-m^2 - i \varepsilon \omega} \\
\hat{S}_{m^2}^\wedge(\omega,k) &= \lim_{\varepsilon \searrow 0} \frac{1}{\omega^2-k^2-m^2 + i \varepsilon \omega}
\end{split}
\eeq
(where the limit~$\varepsilon \searrow 0$ is taken in the distributional sense).
The resulting Fourier transform can be computed explicitly with residues. Indeed, carrying out
the $\omega$-integral by closing the contour in the upper (lower) half plane if~$t<0$ (respectively $t>0$),
we get
\begin{align*}
&S^\wedge_{m^2}(t,x) = \lim_{\varepsilon \searrow 0}
\int_{\R^2} \frac{d\omega\: dk}{(2 \pi)^2}\: \frac{1}{\omega^2-k^2-m^2 + i \varepsilon \omega}\: e^{-i \omega t+ i k x} \\
&= \lim_{\varepsilon \searrow 0}
\int_{\R^2} \frac{d\omega\: dk}{(2 \pi)^2}\: \bigg( \frac{1}{\omega- \sqrt{k^2+m^2} + i \varepsilon} - \frac{1}{\omega+\sqrt{k^2+m^2} + i \varepsilon} \bigg)
\: \frac{e^{-i \omega t+ i k x}}{2\,\sqrt{k^2+m^2}} \\
&= \Theta(t)\: \frac{(-2 \pi i)}{(2 \pi)^2} \int_{-\infty}^\infty \frac{dk}{2\,\sqrt{k^2+m^2}} \:\Big( e^{-i \sqrt{k^2+m^2}\, t}
- e^{i \sqrt{k^2+m^2}\, t} \Big) \: e^{i k x} \\
&= -\Theta(t)\: \frac{1}{\pi} \int_{0}^\infty \frac{dk}{\sqrt{k^2+m^2}} \: \sin \Big(\sqrt{k^2+m^2}\, t \Big)\:\cos(kx) \\
&= \left\{ \begin{array}{c} \omega^2 = k^2+m^2 \\[0.2em] \displaystyle \frac{dk}{\omega} = \frac{d\omega}{k} \end{array} \right\}
= -\Theta(t)\: \frac{1}{\pi} \int_{m}^\infty \frac{d\omega}{\sqrt{\omega^2-m^2}} \: \sin \big(\omega t \big)\:\cos\Big(
\sqrt{\omega^2-m^2} \,x\Big)
\end{align*}
(where~$\Theta$ is the Heaviside function).
The obtained integral is well-defined as an improper Riemann integral. In order to compute it,
it is most convenient to make use of Lorentz invariance, making it possible to restrict attention to the case~$x=0$.
In this case, the Fourier integral can be carried out using Bessel functions
(see~\cite[eq.~10.9.12]{dlmf})
\[ \int_{m}^\infty \frac{d\omega}{\sqrt{\omega^2-m^2}} \: \sin \big(\omega t \big)
= \int_{1}^\infty \frac{d\sigma}{\sqrt{\sigma^2-1}} \: \sin \big(\sigma \,(m t) \big)
= \frac{\pi}{2}\: J_0(mt) \:, \]
giving the explicit formula
\beq \label{Swedge}
S^\wedge_{m^2}(t,x) = -\frac{1}{2}\:\Theta(t)\: \Theta\big(t^2-x^2 \big)\: J_0 \Big( m\, \sqrt{t^2-x^2} \Big) \:.
\eeq
This Green's kernel vanishes unless the point~$(t,x)$ lies in the future light cone centered at the origin.
As a consequence, in the Green's operator~\eqref{greenop} the function~$\phi$ enters
only inside the past light cone centered at~$(t,x)$. This is the reason why~$S^\wedge_{m^2}$ is referred
to as the {\em{retarded Green's operator}}.
Similarly, the Green's kernel~$S^\vee_{m^2}(t,x)$ is computed by
\beq \label{Svee}
S^\vee_{m^2}(t,x) = -\frac{1}{2}\:\Theta(-t)\: \Theta\big(t^2-x^2 \big)\: J_0 \Big( m\, \sqrt{t^2-x^2} \Big) \:,
\eeq
giving rise to the {\em{advanced Green's operator}}~$S^\vee_{m^2}$.

We finally introduce the {\em{fundamental solution}}~$K_{m^2}$ by
\beq \label{Kdef}
\begin{split}
K_{m^2}(t,x) \;&\!\!:= \frac{1}{2 \pi i} \:\big( S_{m^2}^\vee - S_{m^2}^\wedge \big)(t,x) \\
&= -\frac{i}{4 \pi}\:\epsilon(t)\: \Theta\big(t^2-x^2 \big)\: J_0 \Big( m\, \sqrt{t^2-x^2} \Big)
\end{split}
\eeq
(where~$\epsilon$ is the sign function).
Being composed of the difference of the advanced and retarded Green's kernels,
the kernel of the fundamental solution satisfies the homogeneous Klein-Gordon equation,
\beq \label{Khomo}
\big( \partial_t^2 - \partial_x^2 + m^2 \big) \, K_{m^2}(t,x) = 0 \:.
\eeq
For this reason, the fundamental solution can be used to construct solutions
of the Klein-Gordon and wave equations.
The causal fundamental solution has the Fourier representation
\beq \label{Kfourier}
K_{m^2}(t,x) = \int_{\R^2} \frac{d\omega\: dk}{(2 \pi)^2}\: \delta\big( \omega^2-k^2-m^2 \big)\: \epsilon(\omega)
\: e^{-i \omega t+ i k x} \:.
\eeq
Here the fact that the integrand is supported
on the mass shell~$\omega^2+k^2=m^2$ can be understood immediately from
the fact that~$K_{m^2}$ satisfies the Klein-Gordon equation~\eqref{Khomo}.
The detailed form of this integrand can be derived from~\eqref{Kdef} and~\eqref{Sk}
by using the distributional relation
\[ \lim_{\varepsilon \searrow 0} \left( \frac{1}{x - i \varepsilon} - \frac{1}{x + i \varepsilon} \right)
=\, 2 \pi i \: \delta(x) \]
to obtain
\begin{align*}
S_{m^2}^\vee(\omega, k) - S_{m^2}^\wedge(\omega, k)
&= \lim_{\varepsilon \searrow 0} \left[ \frac{1}{\omega^2 - k^{2}-m^{2}-i \varepsilon \omega} -
\frac{1}{\omega^2 - k^{2}-m^{2}+i \varepsilon \omega} \right] \\
&= \lim_{\varepsilon \searrow 0} \left[ \frac{1}{\omega^2 - k^{2}-m^{2}-i \varepsilon} -
\frac{1}{\omega^2 - k^{2}-m^{2}+i \varepsilon} \right] \epsilon(\omega) \\
&= 2 \pi i\, \delta(\omega^2 - k^2 -m^{2})\: \epsilon(q^{0}) \:.
\end{align*}
Alternatively, this relation can also be derived by direct computation of the Fourier integral in~\eqref{Kfourier}.

In the massless case~$m=0$, we obtain the corresponding Green's kernels and
the fundamental solution of the wave equations. Using that~$J_0(0)=1$,
we get the simple formulas
\begin{align}
S^\wedge_0(t,x) &= -\frac{1}{2}\:\Theta(t)\: \Theta\big(t^2-x^2 \big) \label{S0ret} \\
S^\vee_0(t,x) &= -\frac{1}{2}\:\Theta(-t)\: \Theta\big(t^2-x^2 \big) \label{S0adv} \\
K_0(t,x) &= -\frac{i}{4 \pi} \: \epsilon(t) \:\Theta\big(t^2-x^2 \big) \label{K0}
\end{align}
(where~$\epsilon$ is again the sign function).

\section{A Simple Example} \label{secexample} 
The following example is intended to give the reader a first idea of the problem analyzed in this paper.
In particular, the simple arguments presented in this section explain why the answer to the naive question~(A) on page~\pageref{questionA} is no.

Let~$f \in C^\infty_0(\scrM, \C)$ be a compactly supported test function in $1+1$-dimensional
Minkowski space-time~$\scrM$. For notational clarity, we denote points of Minkowski space
in boldface, i.e.\ ${\bf{x}}=({\bf{x}}^0, {\bf{x}}^1)=(t,x)$ and~${\bf{p}}=({\bf{p}}^0, {\bf{p}}^1=k)$.
We again let~$K_0$ be the causal fundamental solution~\eqref{K0}. Then the function
\beq \label{convolve}
\phi({\bf{x}}) := (K_0 f)({\bf{x}}) = \int_\scrM K_0({\bf{x}},{\bf{y}})\: f({\bf{y}})\: d^2{\bf{y}}
\eeq
is a solution of the scalar wave equation which is smooth and has spatially compact support.
Taking the Fourier transform in space and time, the convolution in~\eqref{convolve} becomes a
multiplication in momentum space, i.e.
\beq \label{phimultiply}
\phi({\bf{x}}) = \int_{\R^{2}} \frac{d^{2}{\bf{p}}}{(2 \pi)^{2}}\:
\hat{K}_0({\bf{p}})\: \hat{f}({\bf{p}})\: e^{-i \,\la {\bf{p}}, {\bf{x}} \ra}
\eeq
(where~$\la .,. \ra$ is the Minkowski inner product). Using~\eqref{Kfourier}, the distribution~$\hat{K}_0$
is given by
\[ %\label{hatKdef}
\hat{K}_0({\bf{p}}) = \delta\big( ({\bf{p}}^0)^2 -({\bf{p}}^1)^2 \big)\: \epsilon \big( {\bf{p}}^0 \big) \:. \]
%, the convolution in~\eqref{convolve} becomes a
%multiplication in momentum space, i.e.
%\beq \label{hatphi}
%\hat{\phi}({\bf{p}}) = \hat{K}_0({\bf{p}})\: \hat{f}({\bf{p}}) \:.
%\eeq
We decompose the solution into the components of positive and negative frequencies by setting
\beq \label{phipmmultiply}
\phi_\pm({\bf{x}}) = \int_{\R^{2}} \frac{d^{2}p}{(2 \pi)^{2}}\: \Theta(\pm {\bf{p}}_0)\:
\hat{K}_0({\bf{p}})\: \hat{f}({\bf{p}})\:  e^{-i \,\la {\bf{p}}, {\bf{x}} \ra} %, \qquad \phi({\bf{x}})= \phi_+({\bf{x}})+\phi_-({\bf{x}})
\eeq
and denote their energies by
\[ %\label{Epm}
E\big(\phi_\pm \big) := \frac{1}{2} \int_{-\infty}^\infty \Big( \big|\partial_t \phi_\pm(t,x)\big|^2 + \big|\partial_x \phi_\pm(t,x)\big|^2 \Big)\: dx \:. \]
Clearly, these energies are time independent due to energy conservation.

We now answer question~(A) on page~\pageref{questionA}:
\begin{Prp} For any~$\varepsilon>0$, there is a smooth solution~$\phi({\bf{x}})$ with spatially compact support of
the wave equation in $(1+1)$-dimensional Minkowski space with the property that
\[ \frac{E(\phi_-)}{E(\phi_+)} \leq \varepsilon^2 \:. \]
\end{Prp}
\Proof Given~$f \in C^\infty_0(\scrM)$, in~\eqref{convolve} we consider the family of test functions
\[ %\label{eq:testfunction}
f_\zeta({\bf{x}}) := f({\bf{x}})\: \exp \big(-i \zeta\, ({\bf{x}}^0+{\bf{x}}^1) \big) \:, \]
where~$\zeta$ is a positive parameter. For convenience, the test function~$f$ is chosen such that
$\max_{\R^{2}}(\hat{f})=\hat{f}(0,0)$. 
Taking the Fourier transform, the multiplication by a plane wave translates into a shift of the
argument, i.e.
\[ \hat{f}_\zeta({\bf{p}}) = \hat{f}\big({\bf{p}}^0 - \zeta, {\bf{p}}^1 +\zeta\big) \:. \]
We now consider the corresponding family of solutions~$\phi_\zeta$ in~\eqref{phimultiply}.

\begin{figure}
% \usepackage[usenames,dvipsnames]{pstricks}
% \usepackage{epsfig}
% \usepackage{pst-grad} % For gradients
% \usepackage{pst-plot} % For axes
% \usepackage[space]{grffile} % For spaces in paths
% \usepackage{etoolbox} % For spaces in paths
% \makeatletter % For spaces in paths
% \patchcmd\Gread@eps{\@inputcheck#1 }{\@inputcheck"#1"\relax}{}{}
% \makeatother

\psscalebox{1.0 1.0} % Change this value to rescale the drawing.
{
\begin{pspicture}(-0.5,-1.4393913)(4.282778,1.4393913)
\definecolor{colour0}{rgb}{0.8,0.8,0.8}
\psellipse[linecolor=black, linewidth=0.02, fillstyle=solid,fillcolor=colour0, dimen=outer](1.6875,-0.4643913)(0.3725,0.2)
\psellipse[linecolor=black, linewidth=0.02, fillstyle=solid,fillcolor=colour0, dimen=outer](0.3725,0.8356087)(0.3725,0.2)
\psline[linecolor=black, linewidth=0.04, arrowsize=0.05291667cm 2.0,arrowlength=1.4,arrowinset=0.0]{->}(0.055,-0.46828017)(3.3416667,-0.47494683)
\psline[linecolor=black, linewidth=0.04, arrowsize=0.05291667cm 2.0,arrowlength=1.4,arrowinset=0.0]{->}(1.675,-1.4393913)(1.675,1.4317198)
\psline[linecolor=black, linewidth=0.02](0.043333333,1.1639421)(2.6366668,-1.4293913)
\psline[linecolor=black, linewidth=0.02](0.72333336,-1.4293913)(3.3122222,1.1639421)
\rput[bl](3.182778,-0.9005024){${\bf{p}}^1$}
\rput[bl](1.8444444,1.1639421){${\bf{p}}^0$}
\psline[linecolor=black, linewidth=0.04, arrowsize=0.05291667cm 2.0,arrowlength=1.4,arrowinset=0.0]{->}(1.235,-0.28939128)(0.355,0.5856087)
\rput[bl](2.2177777,-0.3005024){$\hat{f}$}
\rput[bl](0.7927778,0.70949763){$\hat{f}_\zeta$}
\end{pspicture}
}
\caption{Shifting~$\hat{f}_\zeta$ in momentum space. The shaded region indicates the neighborhood around the maximum of $\hat{f}_\zeta$, outside of which $\hat{f}_\zeta$ decays rapidly.}
\label{fig1}
\end{figure}
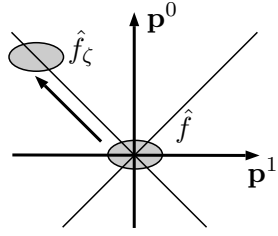

By increasing~$\zeta$, the function~$\hat{f}_\zeta$ is shifted parallel to the light cone
towards higher positive frequencies (see Figure~\ref{fig1}) with $\max_{\R^{2}} \hat{f}_\zeta =\hat{f}(\zeta,-\zeta)$.
As a consequence, the energy~$E(\phi_{\zeta, +})$ of the positive-frequency contribution is bounded from below. 
Furthermore, since~$f({\bf{x}})$ is smooth, its Fourier transform~$\hat{f}$ decays rapidly. As a consequence, $\hat{\phi}_{\zeta, -}$ as well as its energy~$E(\phi_{\zeta,-})$ tend to zero rapidly in~$\zeta$.
Hence
\[ \lim_{\zeta \rightarrow \infty} \frac{E(\phi_{ \zeta,-})}{E(\phi_{\zeta,+})} = 0 \:, \]
concluding the proof.
\QED

This example can be made more quantitative. In order to get a good example
for testing our estimates, we want to choose a compactly supported function of one variable 
whose Fourier transform decays as fast as possible near infinity.
As proven in~\cite[theorem in Section~1.5]{mashreghi}, there is a non-trivial, compactly supported function~$g$
whose Fourier transform is bounded by
\beq \label{gdecay} 
|\hat{g}(k)| \leq \exp \Big( -\frac{|k|}{1 + \log^2 |p|} \Big) \:.
\eeq
This ``almost exponential'' decay near infinity is optimal in the sense that
there is no compactly supported function~$g$ with (see~\cite[theorem in Section~1.1]{mashreghi})
\[ |\hat{g}(k)| \leq \exp \Big( -\frac{|k|}{1 + \log |p|} \Big) \:. \]

We choose
\[ f({\bf{x}}) = g \big({\bf{x}}^0 \big) \: g\big({\bf{x}}^1 \big) \]
with~$g$ satisfying~\eqref{gdecay}.
For this choice of~$g$, we can compute the energies of the corresponding
solutions~$\phi_\zeta$ in~\eqref{phimultiply} and~\eqref{phipmmultiply} as well
as their spatial Fourier transforms~\eqref{fdef1}
explicitly. A straightforward calculation yields
\begin{align}
\big| k\, \hat{\phi}_{\zeta, +}(k) \big| &\leq |k| \: \exp \bigg( -\frac{\big| \zeta-|k| \big|}{1 + \log^2 \big| \zeta- |k| \big| } \bigg)
\label{phip} \\
\big| k\, \hat{\phi}_{\zeta, -}(k) \big| &\leq |k| \: \exp \bigg( -\frac{\zeta+|k|}{1 + \log^2 \big| \zeta+ |k| \big| } \bigg) \\
E(\phi_\zeta) &\sim \zeta^2 \\
E\big( \phi_{\zeta,-} \big) &\lesssim \int_0^\infty \omega^2 \: \exp \bigg( -\frac{2\,(\zeta + \omega)}{1 + \log^2 (\omega+\zeta) } \bigg) \: d\omega \notag \\
&\lesssim \big( 1 + \log^2 \zeta)^3 \: \exp \bigg( -\frac{2 \zeta}{1+\log^2 \zeta} \bigg) \:.
\end{align}
Hence
\beq \label{epszeta}
\varepsilon := \sqrt{\frac{E\big( \phi_{\zeta,-} \big)}{E(\phi_\zeta)}}
\lesssim \frac{( 1 + \log^2 \zeta)^{\frac{3}{2}}}{\zeta} \: \exp \bigg( -\frac{\zeta}{1+\log^2 \zeta} \bigg) \:.
\eeq
Combining the above inequalities, one sees that for fixed~$k$ and small~$\varepsilon$
(i.e.\ for large~$\zeta$), in the above example
the function~$R$ in~\eqref{main} tends to zero in~$\varepsilon$ slightly faster than linearly.
Such a bound of~$\hat{\phi}_\pm(k)$ in terms of~$\varepsilon$ holds as long as the
exponential in~\eqref{phip} is small, i.e.\ as long as~$|k| \lesssim \zeta$.
Inverting~\eqref{epszeta} asymptotically for large~$\zeta$, one finds that~$\zeta \sim -\log \varepsilon$. Therefore,
the interval for~$|k|$ on which our improved estimate applies grows logarithmically in~$\varepsilon$.

These qualitative findings will be reproduced by our estimates. Indeed, we shall see that
for small~$k$ and~$\varepsilon$, the function~$R$ in~\eqref{main} scales like~$R \sim \varepsilon^\frac{2}{3}$
(see Proposition~\ref{prop:series}), which is consistent with the slightly faster than
linear decay in~$\varepsilon$ in the above example. Moreover, the logarithmic growth in~$\varepsilon$ of the
interval~$|k| \in [0, \zeta]$ also appears in our refined estimates
(see for example Proposition~\ref{prpges}, where the region~{\bf{(A)}} is determined by
the inequality~\eqref{kloges} with~$k=\sqrt{2b}$ and~$\lambda$, $a$ and~$b$ as defined by~\eqref{lambdadef}
and~\eqref{abdef} with~$s=1$).

Although the methods used in this example give a good first understanding, 
it seems impossible to use them for proving Theorem~\ref{thmhaupt}. 
One reason is that the methods for analyzing the
decay of Fourier transforms of compactly supported functions
(see~\cite{mashreghi} for a good survey) do not give precise
estimates. Another reason is that in~\eqref{phimultiply} the function~$\hat{f}_\zeta$ is
multiplied by a distribution supported on the mass cone. As a consequence, results
on the decay of two-dimensional Fourier transforms do not seem suitable for analyzing
solutions of the wave equation.

\section{The~\texorpdfstring{$1+1$}{one plus one}-Dimensional Case} \label{sec11dim}
In this section we give a detailed analysis of the properties of solutions to the wave equation 
with spatially compact support in $1+1$-dimensional Minkowski space in the limiting case when
the quotient~$E(\phi_-)/E(\phi_+)$ is small. In particular, we shall derive an upper bound for
the Fourier transform of such solutions for small frequencies.

We consider the Cauchy problem for the
scalar wave equation with smooth initial data supported inside the unit ball~$B_1=(-1,1)$,
\beq \label{cauchy}
\left\{ \begin{array}{c} (\partial_t^2 - \partial_x^2) \phi(t,\vec{x}) = 0 \\[0.3em]
\phi|_{t=0} = \phi_0 \in C^\infty_0(B_1) \:, \qquad \partial_t \phi|_{t=0} = \phi_1 \in C^\infty_0(B_1) \:.
\end{array} \right.
\eeq
We denote the energy of the solution by 
\beq \label{E0}
E(\phi) := \frac{1}{2} \int_{B_1} \Big( \big|\partial_t \phi(0,x) \big|^2 + \big|\partial_x \phi(0,x) \big|^2 \Big)\: dx \:.
\eeq
It is useful to take the Fourier transform of the spatial variable, again using the notation and conventions in~\eqref{fdef1}.
A direct computation yields
\[ \hat{\phi}(t,k) = \hat{\phi}_+(t,k) + \hat{\phi}_-(t,k) \]
with
\beq \label{phipmrep1d}
\hat{\phi}_\pm(t,k)
:= \frac{1}{2} \: e^{\mp i \omega t}\Big( \hat{\phi}_0(k) \pm \frac{i}{\omega}\: \hat{\phi}_1(k) \Big) \:,
\eeq
where $\omega \geq 0$ denotes the absolute value of the frequency, i.e.\
\beq \label{omegadef}
\omega = \omega(k) := |k| \:.
\eeq
The solutions~$\phi_\pm$ can be understood as the components of positive and negative frequency, respectively. This splitting is analogous to the splitting into plus- and minus-functions in~\cite[p. 16]{havin2001uncertainty}.
Using Plancherel's theorem, the energy~\eqref{E0} can also be expressed
as an integral in momentum space.
\begin{Lemma} \label{lemmaE3}
The energy~\eqref{E0} can be written as
\beq \label{E3}
E(\phi) = E(\phi_+) + E(\phi_-) \qquad \text{with} \qquad
E(\phi_\pm) := \int_{-\infty}^\infty \frac{dk}{2 \pi} \;\omega^2\,\big|\hat{\phi}_\pm(k)\big|^2 \:.
\eeq
\end{Lemma}
\Proof A direct computation using Plancherel's theorem yields
\begin{align*}
E(\phi) &= \frac{1}{2} \int_{-\infty}^\infty \frac{dk}{2 \pi} \:\Big( \omega^2\, \big|\hat{\phi}_0(k)\big|^2
+ \big|\hat{\phi}_1(k) \big|^2 \Big) \\ %\label{E1} \\
&= \int_{-\infty}^\infty \frac{dk}{2 \pi} \;\omega^2\,\Big( \big|\hat{\phi}_+(t, k)\big|^2
+ \big|\hat{\phi}_-(t, k)\big|^2 \Big) \:, %\label{E2}
\end{align*}
giving the result.
\QED

We now enter the proof of Theorem~\ref{thmhaupt}
in different versions (see Lemma \ref{constantbound}, Theorem~\ref{maintheoremsimple},
Theorem~\ref{thmmain2} and Corollary~\ref{cormain}).
Our strategy is as follows: We begin with a pointwise bound of the Fourier transform. In order to improve on this result for small frequencies, we expand the Fourier transform in a Taylor series about the origin.
For technical reasons, we consider the contributions of even and odd parity separately.
We successively derive more and more refined estimates for the Taylor coefficients. In the final step, we prove several bounds for the Taylor series in closed form. Our estimates will be presented in increasing level of refinement and, accordingly, in increasing complexity of the proofs.

\subsection{A Pointwise Bound of the Fourier Transform} \label{secpointwise}
We begin with a simple and well-known pointwise bound for the Fourier transform.
It will serve as a reference for the improved bounds for small frequencies to be derived later on.
For our estimates, it is useful to introduce the functions
\[ \hat{h}_\pm(k) := \omega\: \hat{\phi}_\pm(0,k) \]
with~$\omega$ as in~\eqref{omegadef}, where for convenience we evaluated at time $t=0$.
According to Lemma~\ref{lemmaE3}, the energy~$E(\phi_\pm)$ simply is a multiple of
the $L^2$-norm of~$\hat{h}_\pm(k)$ squared.
The following estimates apply similarly to both~$\hat{h}_+$ and~$\hat{h}_-$.
We begin with a pointwise bound.

\begin{Lemma}\label{constantbound} For all~$k \in \R$,
\[ \big|\hat{h}_\pm(k)\big| \leq \sqrt{2 E(\phi)}\:. \]
\end{Lemma}
\Proof According to~\eqref{phipmrep1d},
\[ \big|\hat{h}_\pm(k) \big| = |k\, \hat{\phi}_\pm(k)|
\leq \frac{1}{2} \, \Big( |k\, \hat{\phi}_0(k)| + |\hat{\phi}_1(k)| \Big) 
\leq \frac{1}{\sqrt{2}}\, \Big( |k\, \hat{\phi}_0(k)|^2 + |\hat{\phi}_1(k)|^2 \Big)^\frac{1}{2} 
\:. \]
The obtained Fourier transforms can be estimated pointwise by
\begin{align*}
\big| k \,\hat{\phi}_0(k) \big| &\leq \bigg| \int_{B_1} \partial_x\phi_0(x) \: e^{-i k x}\: dx \bigg|
\leq \int_{B_1} \big|\partial_x\phi_0(x)\big| \: dx \leq \sqrt{2}\: \|\partial_x\phi_0\|_{L^2(B_1)} \\
\big| \hat{\phi}_1(k) \big| &\leq \bigg| \int_{B_1} \phi_1(x) \: e^{-i k x}\: dx \bigg|
\leq \int_{B_1} \big|\phi_1(x)\big| \: dx \leq \sqrt{2}\: \|\phi_1\|_{L^2(B_1)} \:.
\end{align*}
Comparing with~\eqref{E0} evaluated at time~$t=0$ gives the result.
\QED
The goal of the following sections is to improve this estimate of~$|\hat{h}_\pm(k)|$
for small~$k$.

\subsection{Taylor Expansion in Momentum Space} \label{sectaylormom}
Our first step is to expand the initial data~$\hat{\phi}_{0\!/\!1}$ as well as the corresponding
solutions~$\phi_\pm$ of positive and negative frequency in Taylor series about the momentum~$k=0$.
Since the initial data is compactly supported, its Fourier transform is real analytic
(for a proof of this statement see Lemma~\ref{lemma1}).
Therefore, we may expand the initial data in Taylor series,
\beq \label{taylorinit}
\hat{\phi}_0(k) = \sum_{n=0}^\infty \frac{\hat{\phi}^{(n)}_0(0)}{n!}\: k^n \qquad \text{and} \qquad
\hat{\phi}_1(k) = \sum_{n=0}^\infty \frac{\hat{\phi}^{(n)}_1(0)}{n!}\: k^n \:.
\eeq
Using these formulas in~\eqref{phipmrep1d}, we obtain corresponding series expansions
for the solutions~$\hat{\phi}_\pm$ (we evaluate at~$t=0$ and leave out the argument~$t$),
\[ \hat{\phi}_\pm(k)
= \frac{1}{2} \: \Big( \hat{\phi}_0(k) \pm \frac{i}{\omega}\: \hat{\phi}_1(k) \Big)
= \frac{1}{2} \sum_{n=0}^\infty \bigg( \frac{\hat{\phi}^{(n)}_0(0)}{n!} \pm \frac{i}{\omega}\:
\frac{\hat{\phi}^{(n)}_1(0)}{n!} \bigg)\: k^n \:. \]
According to Lemma~\ref{lemmaE3}, the energy is the $L^2$-norm of~$\omega\, \hat{\phi}_\pm(k)$.
Therefore, we multiply by~$\omega$. Using that~$\omega=|k|$, we obtain
\begin{align}
\hat{h}_\pm(k) = \omega\, \hat{\phi}_\pm(k)
&= \frac{1}{2} \sum_{n=0}^\infty \bigg( \omega\, \frac{\hat{\phi}^{(n)}_0(0)}{n!} \pm i\:
\frac{\hat{\phi}^{(n)}_1(0)}{n!} \bigg)\: k^n \notag \\
&= \frac{1}{2} \sum_{n=0}^\infty \bigg( \epsilon(k)\, \frac{\hat{\phi}^{(n)}_0(0)}{n!} \: k^{n+1} \pm i\:
\frac{\hat{\phi}^{(n)}_1(0)}{n!} \: k^n \bigg) \:, \label{twoseries}
\end{align}
where~$\epsilon(k)$ is again the sign function.
This sign function is crucial for what follows.
Its significance becomes clear from the fact that it is responsible
for Hegerfeldt's theorem to hold: Assume that~$\hat{\phi}_-$ vanishes.
Then the series in~\eqref{twoseries} must vanish for all~$k \in \R$.
Hence the coefficient of every power in~$|k|$ must be zero, i.e.
\[  \hat{\phi}^{(0)}_0(0) = 0 \qquad \text{and} \qquad
\epsilon(k)\: \frac{\hat{\phi}^{(n-1)}_0(0)}{(n-1)!}+ i\: \frac{\hat{\phi}^{(n)}_1(0)}{n!} = 0 \quad \text{for all~$n \geq 1$}\:. \]
This equation must hold for both signs of~$k$, i.e.\
\begin{align*}
\frac{\hat{\phi}^{(n-1)}_0(0)}{(n-1)!}+ i\: \frac{\hat{\phi}^{(n)}_1(0)}{n!} &= 0 \qquad \text{for~$k>0$} \\
-\frac{\hat{\phi}^{(n-1)}_0(0)}{(n-1)!}+ i\: \frac{\hat{\phi}^{(n)}_1(0)}{n!} &= 0 \qquad \text{for~$k<0$}\:.
\end{align*}
As a consequence, all the summands in~\eqref{twoseries} must be zero,
implying that the initial data vanishes identically.
This simple argument even makes it possible to quantify Hegerfeldt's theorem.
Indeed, if~$\hat{\phi}_-$ is small, then all its Taylor coefficients are small,
implying that also the initial data must be small.
Clearly, our task is to specify what ``small'' means, and to derive
corresponding estimates.

In preparation of this analysis, we now express the energy of~$\phi_\pm$ in terms of the initial data.
It is useful to decompose the solution with respect to parity, i.e.\ the symmetry
under spatial reflections at the origin. Thus for a function~$\phi(t,x)$ we
introduce the parity decomposition by
\[ \phi(t,x) = \phi^\even(t,x) + \phi^\odd(t,x) \:, \]
where
\[ \phi^\even(t,x) := \frac{1}{2} \Big( \phi(t,x) + \phi(t,-x) \Big) \quad \text{and} \quad
\phi^\odd(t,x) := \frac{1}{2} \Big( \phi(t,x) - \phi(t,-x) \Big)  \:. \]
Since the Fourier transform preserves parity, we obtain similar decompositions in momentum space,
namely
\[ \hat{\phi}^\even(k) = \frac{1}{2} \Big( \hat{\phi}(k) + \hat{\phi}(-k) \Big) \qquad \text{and} \qquad
\hat{\phi}^\odd(k) = \frac{1}{2} \Big( \hat{\phi}(k) - \hat{\phi}(-k) \Big) \:. \]
Having fixed the parity, it clearly suffices to analyze $\hat{\phi}^{\even/\odd}$ for positive~$k$,
implying that~$k=|k|=\omega$. Therefore, it is
unnecessary to distinguish between~$k$ and~$\omega$.
Comparing with~\eqref{twoseries}, we obtain
\beq \label{hatpm}
\hat{h}^\even_\pm(\omega) = \sum_{n=0}^\infty a_n^\even\: \omega^n \qquad \text{and} \qquad
\hat{h}^\odd_\pm(\omega) = \sum_{n=1}^\infty a_n^\odd\: \omega^n \:,
\eeq
where the series coefficients of even and odd parity are given by
\begin{align}
a^\even_{2\ell} &= \pm \frac{i}{2} \: \frac{\hat{\phi}_1^{(2\ell)}(0)}{(2\ell)!} \:,\hspace*{-1.5cm}&
a^\even_{2\ell+1} &= \frac{1}{2}\: \frac{\hat{\phi}_0^{(2\ell)}(0)}{(2\ell)!} \label{even} \\
a^\odd_{2\ell+2} &= \frac{1}{2}\: \frac{\hat{\phi}_0^{(2\ell+1)}(0)}{(2\ell+1)!} \:,\hspace*{-1.5cm}&
a^\odd_{2\ell+1} &= \pm \frac{i}{2}\: \frac{\hat{\phi}_1^{(2\ell+1)}(0)}{(2\ell+1)!} \:. \label{odd}
\end{align}

\begin{Lemma} \label{lemmasplit}
The energy of the positive and negative frequency components of~$\phi$ as given in
Lemma~\ref{lemmaE3}
can be written as
\[ E(\phi_\pm) = E(\phi^\even_\pm) + E(\phi^\odd_\pm) \]
with
\beq \label{sersplit}
\begin{split}
E\big(\phi^\even_\pm \big) &= \frac{1}{\pi} \int_0^\infty 
\bigg| \sum_{n=0}^\infty a^\even_n\, \omega^n \bigg|^2\: d\omega \\
E\big(\phi^\odd_\pm \big) &= \frac{1}{\pi} \int_0^\infty 
\bigg| \sum_{n=1}^\infty a^\odd_n\, \omega^n \bigg|^2\: d\omega \:. 
\end{split}
\eeq
\end{Lemma}
\Proof
Using~\eqref{E3}, we obtain
\begin{align*}
E(\phi_\pm) &= \int_{-\infty}^\infty \frac{dk}{2 \pi} \:\big| \hat{h}_\pm(k) \big|^2 
= \int_0^\infty \frac{d\omega}{2 \pi} \:\Big( \big| \hat{h}_\pm(\omega) \big|^2 + \big| \hat{h}_\pm(-\omega) \big|^2 \Big) \\
&= \frac{1}{4 \pi} \int_0^\infty 
\Big( \big| \hat{h}_\pm(\omega) + \hat{h}_\pm(-\omega) \big|^2 + \big| \hat{h}_\pm(\omega) - \hat{h}_\pm(-\omega) \big|^2 \Big)
\: d\omega \:.
\end{align*}
The two summands in the integrand are the even and odd parity components, respectively.
Computing them using~\eqref{hatpm} gives the result.
\QED

\subsection{Simple Estimates of the Taylor Coefficients} \label{secsimpes}
The following estimates apply to both series in~\eqref{sersplit}
in the same way. For notational convenience, the superscript~$\eo$
stands for either ``even'' or ``odd.''
Thus we write the series in~\eqref{sersplit} as
\beq \label{fdef}
\hat{h}_\pm^\eo(\omega) := \sum_{n=0}^\infty a_n^\eo \,\omega^n \::\: \R^+ \rightarrow \C \:,
\eeq
where we set~$a_0^\odd=0$.
Our goal is to estimate the functions~$\hat{h}_\pm^\eo(\omega)$ for low frequencies.
Before entering this analysis, we point out that, according to~\eqref{even} and~\eqref{odd},
the coefficients~$a_n^\eo$ differ in the cases~$+$ and~$-$ only by signs.
Therefore, whenever we estimate the absolute values of these coefficients, the distinction between
the cases~$+$ and~$-$ becomes irrelevant. Moreover, from~\eqref{even} and~\eqref{odd}
one sees that the series involving the absolute values of the coefficients bounds the initial data
in the sense that
\[ 2\, \big| \hat{h}^\eo_\pm(k) \big| \leq \big| k\, \hat{\phi}^\eo_0(k) \big| + \big| \hat{\phi}^\eo_1(k) \big|
\leq \sum_{n=0}^\infty \big|a_n^\eo\big| \,\omega^n \:. \]
These inequalities will be crucial for the following estimates.

We begin with a simple estimate of each coefficient of the series expansion, which is
based on Lemma~\ref{lemma1}.
\begin{Prp} \label{prp1}
The coefficients in the power series~\eqref{fdef} are bounded by
\[ |a^\eo_n| \leq \frac{\sqrt{E(\phi^\eo)}}{n!} \:. \]
\end{Prp} 
\Proof
Using the result of Lemma~\ref{lemma1} in~\eqref{even} and~\eqref{odd},
one finds that the coefficients~$a^\eo_n$ are bounded by
\begin{align*}
\big|a^\even_{2\ell} \big| &\leq \frac{1}{\sqrt{2}}\: \frac{1}{(2\ell)!} \: \|\phi^\even_1\|_{L^2(B_1)}\:, &
\big|a^\even_{2\ell+1} \big| &\leq \frac{1}{\sqrt{2}}\: \frac{1}{(2\ell+1)!}
\: \|\partial_x \phi^\even_0\|_{L^2(B_1)} \\
\big|b^\odd_{2\ell+2} \big| &\leq \frac{1}{\sqrt{2}}\: \frac{1}{(2\ell+2)!} 
\: \|\partial_x \phi^\odd_0\|_{L^2(B_1)}\:,&
\big|b^\odd_{2\ell+1} \big| &\leq \frac{1}{\sqrt{2}}\: \frac{1}{(2\ell+1)!}
\: \|\phi^\odd_1\|_{L^2(B_1)} \:.
\end{align*}
We thus obtain the simple bound in terms of the energy
\begin{align*}
|a^\eo_n| &\leq \frac{1}{n!}\: \frac{1}{\sqrt{2}}\: \max \Big\{ \|\partial_x\phi^\eo_0\|_{L^2(B_1)},\:
 \|\phi^\eo_1\|_{L^2(B_1)} \Big\} \\
&\leq \frac{1}{n!}\: \frac{1}{\sqrt{2}}\: \sqrt{ \|\partial_x\phi^\eo_0\|_{L^2(B_1)}^2 + \|\phi^\eo_1\|_{L^2(B_1)}^2 } 
= \frac{\sqrt{E(\phi^\eo)}}{n!} \:.
\end{align*}
This concludes the proof.
\QED

\subsection{Estimates of the Highest Coefficient of a Polynomial} \label{sechighcoeff}
In Proposition~\ref{prp1} the Taylor coefficients were estimated in terms of the
total energy~$E(\phi^\eo)$ of the wave. However, it was not taken into account
that the corresponding Taylor series describes the component of positive or negative
frequency only (see~\eqref{hatpm}). More specifically, we consider the
situation when the energy of the negative-frequency component is much smaller than the
total energy,
\[ E(\phi^\eo_-) \ll E(\phi^\eo) \:. \]
Choosing the plus sign in~\eqref{hatpm}, we are interested in upper bounds of the
Taylor coefficients in~\eqref{fdef} which tend to zero if~$E(\phi^\eo_-)$ tends to zero
for fixed~$E(\phi^\eo)$. In order to derive these refined estimates, we use the
following strategy, which is similar to that used by Tao to prove a version of Hardy's uncertainty principle in~\cite[Section 2.6.2., p.360]{tao2010epsilon} . We decompose the Taylor series into a Taylor polynomial
of degree~$N$ and the remainder term,
\beq \label{decompose}
\hat{h}_\pm^\eo = \hat{h}^\eo_N + R^\eo_N \qquad \text{with} \qquad
\hat{h}^\eo_N(\omega) := \sum_{n=0}^N a^\eo_n \, \omega^n \:,\quad
R^\eo_N(\omega) := \sum_{n=N+1}^\infty a^\eo_n \, \omega^n \:.
\eeq
We first show that if the Taylor polynomial has small $L^2$-norm on an interval~$[0, \omega_1]$,
then its highest coefficient must also be small. This statement is quantified in the following lemma
using properties of the Legendre polynomials.
Combining this statement with an $L^2$-estimate of the remainder term (see Lemma~\ref{lemmaremainder}
in the next section), we shall obtain the refined estimates of each Taylor coefficient in Proposition~\ref{prpanes}.

\begin{Lemma} \label{lemmapoly} Let~${\mathcal{P}}(\omega)$ be a real polynomial of degree at most~$N$
with~$N \in \N_0$,
\[ {\mathcal{P}}(\omega) = a_0 + a_1\, \omega + \cdots + a_N\: \omega^N \:. \]
Then for any~$\omega_1>0$, the highest coefficient of~${\mathcal{P}}$
satisfies the following inequalities:
\begin{align}
|a_N| &\leq \frac{1}{\sqrt{\omega_1}}\: \sqrt{\frac{2}{\pi}}\:
\bigg( \frac{4}{\omega_1} \bigg)^{N}\: \|{\mathcal{P}}\|_{L^2([0,\omega_1])}\: \Big( 1 + \O \Big( \frac{1}{N} \Big) \Big) \label{ap1} \\
&\leq \frac{1}{\sqrt{\omega_1}}\: \bigg( \frac{4}{\omega_1} \bigg)^N\:
\|{\mathcal{P}}\|_{L^2([0,\omega_1])} \:. \label{ap2}
\end{align}
\end{Lemma}
\Proof For notational simplicity, we arrange by a rescaling that~$\|{\mathcal{P}}\|_{L^2([0,\omega_1])} = 1$.
We make use of the fact that the Legendre polynomials~$P_n$ are orthogonal in~$L^2([-1,1])$.
More precisely, for all~$n,n' \in \N_0$ (see~\cite[Table~18.3.1]{dlmf})
\[ \int_{-1}^1 P_n(x)\: P_{n'}(x) = \frac{2}{2n+1}\: \delta_{n,n'} \:. \]
Combining this orthogonality with the fact that the 
Legendre polynomials~$P_0, \ldots, P_{N-1}$ are a basis of the polynomials of
degree at most~$N-1$, we conclude that the Legendre polynomial~$P_N$ is orthogonal to all polynomials of degree smaller than~$N$.
It follows that
\[ \int_{0}^{\omega_1} {\mathcal{P}}(\omega)\: P_{N}\Big(\frac{2\omega}{\omega_1}- 1 \Big)
\:d\omega = \int_0^{\omega_1} a_N \:\omega^N\: P_{N}\Big(\frac{2\omega}{\omega_1}- 1 \Big)\:d\omega \:. \]
This makes it possible to compute the coefficient~$a_N$ by
\beq \label{integrals}
a_N = \frac{1}{c_N} \int_{0}^{\omega_1} {\mathcal{P}}(\omega)\: P_{N}\Big(\frac{2\omega}{\omega_1}- 1 \Big)
\:d\omega \quad \text{with} \quad
c_N := \int_0^{\omega_1} \omega^N\: P_{N}\Big(\frac{2\omega}{\omega_1}- 1 \Big)\:d\omega \:.
\eeq
The first integral can be estimated with the help of the Schwarz inequality by
\begin{align}
&\bigg| \int_{0}^{\omega_1} {\mathcal{P}}(\omega)\: P_{N}\Big(\frac{2\omega}{\omega_1}- 1 \Big)
\:d\omega \bigg|
\leq \|{\mathcal{P}}\|_{L^2([0,\omega_1], d\omega)}
\bigg( \int_{0}^{\omega_1} \Big| P_{N}\Big(\frac{2\omega}{\omega_1}- 1 \Big) \Big|^2\:d\omega
\bigg)^\frac{1}{2} \notag \\
&\leq \sqrt{\frac{\omega_1}{2}} \:\|P_{N}\|_{L^2([-1,1])}
= \sqrt{\frac{\omega_1}{2}} \: \frac{\sqrt{2}}{\sqrt{2N+1}}
= \frac{\sqrt{\omega_1}}{\sqrt{2N+1}} \:. \label{intP}
\end{align}

The second integral in~\eqref{integrals}, on the other hand, can be computed
explicitly. First, introducing the integration variable~$x=2\omega/\omega_1 - 1$,
we find that
\begin{align*}
c_N &= \frac{\omega_1}{2} \int_{-1}^1 \Big( \frac{\omega_1\, (x+1)}{2}\Big)^N\: P_{N}(x)\:dx
= \Big( \frac{\omega_1}{2} \Big)^{N+1} \int_{-1}^1 (x+1)^N\: P_{N}(x)\:dx \\
&= \Big( \frac{\omega_1}{2} \Big)^{N+1} \int_{-1}^1 x^N\: P_{N}(x)\:dx
= \Big( \frac{\omega_1}{2} \Big)^{N+1} \:2 \int_0^1 x^N\: P_{N}(x)\:dx\:,
\end{align*}
where in the last line we again used that~$P_N$ is orthogonal to all
polynomials of degree smaller than~$N$.
We now employ the relations (see~\cite[eqs~18.17.38 and 18.17.39]{dlmf})
together with the Stirling formula (see~\cite[eq.~5.11.3 with leading term]{dlmf}),
\begin{align*}
\int_{0}^{1} &P_{2n}\left(x\right)x^{2n}\mathrm{d}x
= \int_{0}^{1}P_{2n}\left(x\right)x^{z-1}\mathrm{d}x \Big|_{z=2n+1}
=\frac{(-1)^{n}{\left(\frac{1}{2}-\frac{1}{2}z\right)_{n}}}{2{\left(\frac{1}{2}z\right)_{n+1}}} \Big|_{z=2n+1} \\
&=\frac{(-1)^{n}{\left(-n\right)_{n}}}{2{\left(n+\frac{1}{2}\right)_{n+1}}} 
=\frac{n!}{2\, (n+\frac{1}{2}) (n+\frac{3}{2}) \cdots (2n+\frac{1}{2})}
%\\ &=\frac{n!\: \Gamma(n+\frac{1}{2})}{2\, \Gamma(2n+\frac{3}{2})} 
%= \frac{n!\: (2n-1)!!\, 2^{2n+1}}{2\, 2^n (4n+1)!!} 
= \frac{n!\, 2^n\: (2n-1)!!}{(4n+1)!!} \\
&= \frac{\sqrt{\pi}}{2}\: \frac{1}{\sqrt{2n}\: 2^{2n}} \Big( 1 + \O \Big( \frac{1}{n} \Big) \Big) \\
%= \frac{1}{2n} + \O \Big( \frac{1}{n^2} \Big) \\
\int_{0}^{1}& P_{2n+1}\left(x\right)x^{2n+1}\mathrm{d}x =
\int_{0}^{1}P_{2n+1}\left(x\right)x^{z-1}\mathrm{d}x \Big|_{z=2n+2}
=\frac{(-1)^{n}{\left(1-\frac{1}{2}z\right)_{n}}}{2{\left(\frac{1}{2}+\frac{1}{2}z\right)_{n+1}}}  \Big|_{z=2n+2} \\
&= \frac{(-1)^{n}{\left( -n \right)_{n}}}{2{\left(n+\frac{3}{2}\right)_{n+1}}} 
=\frac{n!}{2 \,(n+\frac{3}{2}) (n+\frac{5}{2}) \cdots (2n+\frac{3}{2})}
= \frac{n!\:2^n\: (2n+1)!!}{(4n+3)!!} \\
&= \frac{\sqrt{\pi}}{2}\: \frac{1}{\sqrt{2n+1}\: 2^{2n+1}} \Big( 1 + \O \Big( \frac{1}{n} \Big) \Big) \:.
\end{align*}
We thus obtain the estimate
\[ c_N = \sqrt{\pi}\: \Big( \frac{\omega_1}{2} \Big)^{N+1}\: \frac{1}{\sqrt{N}\: 2^N} 
\: \bigg( 1 + \O \Big( \frac{1}{N} \Big) \bigg) \:. \]
Employing the above estimates in~\eqref{integrals} gives~\eqref{ap1}.

Clearly, the relation~\eqref{ap1} implies that~\eqref{ap2} holds for large~$N$.
In order to also verify~\eqref{ap2} for small~$N$, one can estimate the
above combinatorial factors directly to obtain
\begin{align*}
\int_{0}^{1} P_{2n}\left(x\right)x^{2n}\mathrm{d}x
&\geq \frac{1}{\sqrt{2\,(2n)+1}\: 2^{2n}} \\
\int_{0}^{1} P_{2n+1}\left(x\right)x^{2n+1}\mathrm{d}x
&\geq \frac{1}{\sqrt{2\,(2n+1)+1}\: 2^{2n+1}} \:.
\end{align*}
As a consequence,
\[ c_N \geq \Big( \frac{\omega_1}{2} \Big)^{N+1}\: \frac{1}{\sqrt{N+1}\: 2^N} \:. \]
Using this estimate together with~\eqref{intP} in~\eqref{integrals} gives~\eqref{ap2}.
\QED

\subsection{Smallness of the Taylor Coefficients} \label{secsmalltaylor}
We next estimate the $L^2$-norm of the remainder term in~\eqref{decompose}
on an interval~$[0, \omega_1]$.
\begin{Lemma} \label{lemmaremainder}
Given~$\varepsilon \in [0, 1]$ and~$N \in \N_0$, we choose
\beq \label{omegamax}
\omega_1 = \Big( \varepsilon^2\: (N+1)!^2\, (2N+3) \Big)^{\frac{1}{2N+3}} \:.
\eeq
Then the remainder term in~\eqref{decompose} is bounded on~$[0, \omega_1]$ by
\[ \|R^\eo_N(\omega)\|_{L^2([0, \omega_1])} \leq 4 \varepsilon\; \sqrt{E(\phi^\eo)} \:. \]
\end{Lemma}
\Proof Applying Proposition~\ref{prp1}, we can estimate the remainder by
\begin{align}
|R^\eo_N(\omega)| &\leq \sum_{n=N+1}^\infty \frac{\omega^n}{n!} \: \sqrt{E(\phi^\eo)} \notag \\
&= \frac{\omega^{N+1}}{(N+1)!} \:\Big( 1 + \frac{\omega}{N+2} + \frac{\omega^2}{(N+2)(N+3)} + \cdots \Big)
\: \sqrt{E(\phi^\eo)} \notag \\
&\leq c(\omega)\: \frac{\omega^{N+1}}{(N+1)!} \:\sqrt{E(\phi^\eo)}\qquad \text{with} \qquad c(\omega) := \sum_{n=0}^\infty \Big( \frac{\omega}{N+2} \Big)^n \:.
\label{cdef}
\end{align}
Choosing~$\omega_1$ according to~\eqref{omegamax}, we know that
for all~$\omega \in [0,\omega_1]$,
\[ \frac{\omega}{N+2} \leq \frac{\omega_1}{N+2} \leq
\frac{\big( (N+1)!^2\, (2N+3) \big)^{\frac{1}{2N+3}}}{N+2} \leq \frac{3}{4}\:, \]
where the last inequality is verified by direct inspection and using the Stirling formula.
Therefore, the geometric series in~\eqref{cdef} converges and is bounded by four,
\[ |R^\eo_N(\omega)| \leq 4\: \frac{\omega^{N+1}}{(N+1)!}\: \sqrt{E(\phi^\eo)} \:. \]
Using this pointwise bound, the $L^2$-norm can be estimated by
\[ \|R^\eo_N(\omega)\|^2_{L^2([0, \omega_1]} \leq 16\: E(\phi^\eo)
\int_0^{\omega_1} \frac{\omega^{2N+2}}{(N+1)!^2}\: d\omega
\leq \frac{16\: E(\phi^\eo)}{(N+1)!^2\, (2N+3)} \:\omega_1^{2N+3} \:, \] %\label{Res} \]
giving the result.
\QED

%\[ \frac{1}{6}\: \varepsilon^{\frac{2}{2N+3}} \: (2N+3) \leq \omega_1 \leq \frac{1}{2}\: \varepsilon^{\frac{2}{2N+3}} \: (2N+3) \:. \]

\begin{Prp} \label{prpanes}
Assume that
\[ E(\phi^\eo_-) \leq \varepsilon^2\: E(\phi^\eo) \:. \]
Then the Taylor coefficients in~\eqref{fdef} are bounded for all~$n \in \N_0$ by
\[ |a^\eo_n| \leq \frac{6}{\sqrt{2n+1}}\: \frac{4^n}{n!}\: \varepsilon^{\frac{2}{2n+3}}
\: \sqrt{E(\phi^\eo)}\:. \]
\end{Prp}
\Proof Given~$N \in \N_0$, we choose~$\omega_1$ as in~\eqref{omegamax}. Then
the $L^2$-norm of the remainder is bounded according to Lemma~\ref{lemmaremainder}.
Combining this fact with Lemma~\ref{lemmasplit}, we obtain
\begin{align*}
\|\hat{h}^\eo_N(\omega)\|_{L^2([0, \omega_1])} &= \big\| \hat{h}_\pm^\eo - R_N^\eo \big\|_{L^2([0, \omega_1])}
\leq \big\| \hat{h}_\pm^\eo \big\|_{L^2([0, \omega_1])} + \big\| R_N^\eo \big\|_{L^2([0, \omega_1])} \\
&\leq \big\| \hat{h}_\pm^\eo \big\|_{L^2([0, \infty))} + \big\| R_N^\eo \big\|_{L^2([0, \omega_1])}
\leq \sqrt{\pi\, E(\phi^\eo_-)} + \|R_N^\eo\|_{L^2([0, \omega_1])} \\
&\leq \varepsilon\, \sqrt{\pi\, E(\phi^\eo)} + 4 \varepsilon\; \sqrt{E(\phi^\eo)}
\leq 6 \varepsilon \, \sqrt{E(\phi^\eo)} \:.
\end{align*}
Applying Lemma~\ref{lemmapoly} to the polynomial~$\hat{h}^\eo_N$ gives the bound
\begin{align*}
|a^\eo_N| &\leq \frac{1}{\sqrt{\omega_1}}\: \Big( \frac{4}{\omega_1} \Big)^N\:  6 \varepsilon \, \sqrt{E(\phi^\eo)} \\
&= \varepsilon^{\frac{2}{2N+3}} \: 4^N\: (N+1)!^{-\frac{2N+1}{2N+3}}\: (2N+3)^{-\frac{2N+1}{4N+6}}\:
6\, \sqrt{E(\phi^\eo)} \:.
\end{align*}
The result follows asymptotically from the Stirling formula
and for small values of~$n$ directly by numerical evaluation.
\QED

\subsection{Smallness of the Initial Data} \label{secsmallinit}
In Proposition~\ref{prpanes} we estimated all the Taylor coefficients~$a^\eo_n$.
According to~\eqref{even} and~\eqref{odd} this also gives control of all the Taylor
coefficients of the initial data~$\hat{\phi}_0$ and~$\hat{\phi}_1$. We thus obtain the following result.
\begin{Prp} \label{prop:series} Assume that the energy of the negative-frequency component is
boun\-ded in terms of the total energy by 
\[ E(\phi^\eo_-) \leq \varepsilon^2\: E(\phi^\eo) \:. \]
Then the even and odd components of the initial data in momentum space are bounded pointwise for all~$\omega \in \R^+$ by
\[ %\label{initestimate}
2\,\big|\hat{h}_\pm^\eo(\omega)\big|\leq \big| \omega \,\hat{\phi}^\eo_0(\omega) \big| + \big| \hat{\phi}^\eo_1(\omega) \big| \leq 12\: \sqrt{E(\phi^\eo)} \;
\big( 4\omega \big)^{-\frac{3}{2}}\: g\big(\omega, \varepsilon \big) \:, \]
where~$g$ is the series
\beq \label{gseries}
g(\omega, \varepsilon) := \sum_{n=0}^\infty 
\frac{1}{\sqrt{2n+1}}\: \frac{(4 \omega)^{n+\frac{3}{2}}}{n!}\: \varepsilon^{\frac{2}{2n+3}}
\eeq
\end{Prp}
\Proof According to~\eqref{taylorinit},
\[ \big| k \,\hat{\phi}^\eo_0(k) \big| + \big| \hat{\phi}^\eo_1(k) \big| \leq
\sum_{n=0}^\infty \bigg( \frac{|(\hat{\phi}^\eo_0)^{(n)}(0)|}{n!}\: |k|^{n+1} + \frac{|(\hat{\phi}^\eo_1)^{(n)}(0)|}{n!}\: |k|^n \bigg) \:. \]
Using~\eqref{even} and~\eqref{odd}, one verifies both for the even and odd components that
\[ \big| k \,\hat{\phi}^\eo_0(k) \big| + \big| \hat{\phi}^\eo_1(k) \big| \leq
2 \sum_{n=0}^\infty |a_n^\eo|\: |k|^n\:. \]
Applying the estimate of Proposition~\ref{prpanes} gives the result.
\QED 
Before studying the series~\eqref{gseries} in detail and deriving bounds in closed form,
we explain how to derive corresponding estimates for both parity components
together (i.e.\ without decomposing into even and odd components).

\begin{Thm} \label{fullsolution}
Assume that the energy of the negative-frequency component is bounded in terms of the total energy by
\[ E(\phi_-) \leq \varepsilon^2\: E(\phi) \:. \]
Then we have the pointwise bound
\[ \big|\hat{h}_\pm(k)\big|\leq 12\,\sqrt{E(\phi)} \:(4\omega)^{-\frac{3}{2}}\:g(\omega,\varepsilon) \:. \]
\end{Thm}
\Proof Clearly, we may assume that both~$E(\phi^{\odd})$ and~$E(\phi^{\even})$ are non-zero,
because otherwise the result follows immediately from Proposition~\ref{prop:series}.
Since~$g$ is monotone increasing in~$\varepsilon$, we may assume that
\beq \label{ineedthislabel}
E(\phi_-)=\varepsilon^2\: E(\phi) \:.
\eeq
Setting~$\delta = E(\phi^{\odd})/E(\phi) \in (0,1)$ and using Lemmas \ref{lemmasplit} and~\ref{lemmaE3},
we find that
\beq \label{deltarel}
E(\phi^{\odd}) = \delta\: E(\phi) \:, \qquad E(\phi^{\even}) = (1-\delta)\: E(\phi) \:.
\eeq
Moreover, we introduce parameters~$\varepsilon_\eo \geq 0$ such that
\beq \label{epsoddeven}
E(\phi^{\odd}_-) = \varepsilon_{\odd}^2\: E(\phi^{\odd}) \:,\qquad
E(\phi^{\even}_-) = \varepsilon_{\even}^2\: E(\phi^{\even}) \:.
\eeq
It follows that
\begin{align*}
\varepsilon^2 E(\phi)&=E(\phi_-) =E(\phi^{\odd}_-) + E(\phi^{\even}_-) \notag \\
    &=\varepsilon^2_{\odd}\,E(\phi^{\odd})+ \varepsilon^2_{\even}\,E(\phi^{\even})
    = \big(\varepsilon^2_{\odd}\,\delta + \varepsilon^2_{\even}\,(1-\delta) \big)\, E(\phi) \:.
\end{align*}
Solving for $\varepsilon_{\even}$ gives
\[ \varepsilon_{\even}= \sqrt{\frac{\varepsilon^2-\varepsilon_{\odd}^2\,\delta}{1-\delta}} \:. \]
This relation shows that~$\varepsilon_{\odd} \geq \varepsilon$ implies $\varepsilon_{\even} \leq \varepsilon$
and vice versa. Therefore, we may assume without loss of generality that~$\varepsilon_{\even} \leq \varepsilon$ 
and~$\varepsilon_{\odd}\geq\varepsilon$ (otherwise we repeat the following argument withe odd and even components interchanged).

Next, it is straightforward to see that 
\[ |\hat{h}_\pm(k)|^2 = \big(|\hat{h}_\pm^{\odd}(k)+ \hat{h}_\pm^{\even}(k)| \big)^2
\leq \big( |\hat{h}_\pm^{\odd}(k)|+ |\hat{h}_\pm^{\even}(k)| \big)^2\leq 2\,
\big(|\hat{h}_\pm^{\odd}(k)|^2+ |\hat{h}_\pm^{\even}(k)|^2 \big) \:. \]
Applying Proposition~\ref{prop:series}, we obtain
\[ \big|\hat{h}_\pm(k)\big|^2\leq \frac{288}{(4\omega)^3}\: \Big( \delta\: g^2(\omega,\varepsilon_{\odd})
+ (1-\delta)\: g^2(\omega, \varepsilon_{\even}) \Big)\, E(\phi) \:. \]
Since~$g$ is monotone increasing in the argument~$\varepsilon$, we may replace~$\varepsilon_\even$
by~$\varepsilon$. Moreover, combining~\eqref{ineedthislabel} with~\eqref{deltarel} and~\eqref{epsoddeven},
one sees that~$\delta\leq \varepsilon^2/\varepsilon_{\odd}^2$. We thus obtain
\begin{equation}\label{nextequation}
\big|\hat{h}_\pm(k)\big|^2\leq \frac{288}{(4\omega)^3}\left( g^2(\omega,\varepsilon_{\odd}) \:\frac{\varepsilon^2}{\varepsilon_{\odd}^2} + g^2\left(\omega,\varepsilon\right)\right)E(\phi).
\end{equation}
Finally, the computation
\begin{align*}
&\frac{\partial}{\partial \varepsilon_{\odd}} \bigg(
g^2(\omega,\varepsilon_{\odd}) \:\frac{\varepsilon^2}{\varepsilon_{\odd}^2} \bigg) 
= \frac{ 2 \varepsilon^2}{\varepsilon_{\odd}^3}\: g(\omega,\varepsilon_{\odd}) \:
\bigg(\varepsilon_{\odd} \:\frac{\partial g(\omega,\varepsilon_{\odd})}{\partial \varepsilon_{\odd}}-g(\omega,\varepsilon_{\odd})\bigg) \\
&= \frac{ 2 \varepsilon^2}{\varepsilon_{\odd}^3}\: g(\omega,\varepsilon_{\odd}) \:
\bigg( \sum_{n=0}^\infty 
\frac{1}{\sqrt{2n+1}}\: \frac{(4 \omega)^{n+\frac{3}{2}}}{n!}\: \varepsilon_{\odd}^{\frac{2}{2n+3}} \:\Big(\frac{2}{2n+3}-1\Big) \bigg) <0
\end{align*}
allows us to set $\varepsilon_{\odd}=\varepsilon$ in~\eqref{nextequation}. This gives the result. 
\QED

\subsection{A First Version of the Main Theorem} \label{secsimple}
The remaining task is to estimate the series~$g(\omega, \varepsilon)$ in~\eqref{gseries},
which we also write as
\begin{equation}\label{eq:series}
R(\omega, \varepsilon) := (4\omega)^{-\frac{3}{2}} \:g(\omega, \varepsilon)= \sum_{n=0}^\infty 
\frac{1}{\sqrt{2n+1}}\: \frac{(4 \omega)^{n}}{n!}\: \varepsilon^{\frac{2}{2n+3}}
\end{equation}
We now prove the first version of our main result. 
\begin{Thm}\label{maintheoremsimple}
Assume that the energy of the negative-frequency component is boun\-ded in terms of the total energy by 
\[ E(\phi^\eo_-) < \varepsilon^2\: E(\phi^\eo) \:. \]
Then the even and odd components of the initial data in momentum space are bounded pointwise for all~$k \in \R$ by
\beq \label{initestimate2}
2\, \big|\hat{h}_\pm^\eo(k)\big|\leq \big| k \,\hat{\phi}^\eo_0(k) \big| + \big| \hat{\phi}^\eo_1(k) \big| \leq 6^{\frac{3}{2}}\: \frac{\sqrt{E(\phi^\eo)}}{\sqrt{2e\,|\log \varepsilon|}} \:e^{4\omega} \:.
\eeq
\end{Thm}
\Proof We estimate the series in~\eqref{eq:series} by
\begin{align*}
&\sum_{n=0}^\infty 
\frac{1}{\sqrt{2n+1}}\: \frac{(4 \omega)^{n}}{n!}\: \varepsilon^{\frac{2}{2n+3}} \leq \sqrt{\frac{3}{2}} \:\sum_{n=0}^\infty 
\sqrt{\frac{2}{2n+3}}\: \frac{(4 \omega)^{n}}{n!}\:  \varepsilon^{\frac{2}{2n+3}} \\
&\leq \sqrt{\frac{3}{2}} \max_{n\in [0,\infty)} \bigg[\sqrt{\frac{2}{2n+3}}\:\varepsilon^{\frac{2}{2n+3}} \bigg]
\;\sum_{n=0}^\infty\frac{(4 \omega)^{n}}{n!} \leq
\sqrt{\frac{3}{2}} \: \sup_{x \in \R^+} \Big[ x\: e^{x^2 \log \varepsilon} \Big] \:e^{4\omega}\:,
\end{align*}
where in the last step we set~$x=\sqrt{2/(2n+3)}$. In order to estimate the last supremum, we
set~$y = \sqrt{-\log \varepsilon} x$,
\[ \sup_{x \in \R^+} \Big[ x\: e^{x^2 \log \varepsilon} \Big]
= \frac{1}{\sqrt{-\log \varepsilon}}\: \sup_{y \in \R^+} y\: e^{-y^2} = \frac{1}{\sqrt{2 e \:|\log \varepsilon|}} \:, \]
where we used that the function~$y e^{-y^2}$ attains its maximum at~$y=\sqrt{2}$.
Combining this estimate with the result from Proposition~\ref{prop:series} gives the result. 
\QED    
Note that the above estimate is an improvement over Lemma \ref{constantbound} as long as
\[ \frac{6^{\frac{3}{2}}\,e^{4\omega}}{\sqrt{4e \,|\log \varepsilon|}}\leq 1 \:. \]

A straightforward calculation gives the following corollary:
\begin{Corollary} \label{corL12}
Assume that the energy of the negative-frequency component is boun\-ded in terms of the total energy by 
\[ E(\phi^\eo_-) \leq \varepsilon^2\: E(\phi^\eo) \:. \]
Then the $L^1$- and $L^2$-norms of the even and odd components of the initial data are bounded 
in momentum space for small frequencies
\beq \label{omegamaxcor}
\omega\leq \omega_{\max}(\varepsilon) := \frac{1}{4}\:
\log\left(\frac{\sqrt{2e\,|\log \varepsilon|}}{6^{\frac{3}{2}}}\right)
\eeq
by
\[ \big\| \hat{h}_\pm^\eo(k)\big\|_{L^1([0,\omega_{\max}(\varepsilon)])} \leq \frac{1}{8} \:\sqrt{E(\phi^\eo)}  \quad \text{ and } \quad \big\| \hat{h}_\pm^\eo(k)\big\|^2_{L^2([0,\omega_{\max}(\varepsilon)])} \leq \frac{1}{32} \:E(\phi^\eo) \:. \]
\end{Corollary} \noindent
From Lemma~\ref{lemmaE3} we know that the $L^2$-norm of~$\hat{h}_\pm^\eo$ on the whole interval~$[0, \infty)$
gives a multiple of the total energy. We thus obtain
\[ \sum_{\pm} \int_{\omega_{\max}(\varepsilon)}^\infty \frac{dk}{2 \pi}\: \omega^2\:\big|\phi^\eo_\pm(k) \big|^2 \geq \Big( 1
- \frac{1}{32\, \pi} \Big) \: E(\phi^\eo) \:. \]
This inequality quantifies that the wave must have a significant high-energy contribution.
Even more, as the function~$\omega_{\max}(\varepsilon)$ is monotone decreasing in $\varepsilon\in (0,1]$
and tends to infinity as~$\varepsilon \searrow 0$, we see that in this limiting case,
the wave must have large contributions of higher and higher frequency.

We now give a less quantitative version of this result, which might be interesting in the context of a Littlewood-Paley decomposition.
\begin{Corollary}
For every compact frequency range $[\omega_0,\omega_1] \subset \R$, every time~$t_0 \in \R$
and every radius~$r$, there is a constant $C<1$ such that the a-priori estimate
\begin{equation*}
    E(\pi_{[\omega_0,\omega_1]}\phi) \leq C E(\phi)
\end{equation*}
holds for every smooth solution to the $1+1$-dimensional wave equation with
\[ \supp \phi(t_0,.) \subset B_r \:. \]
Here~$\pi_{[\omega_0,\omega_1]}\phi$ is the projection of the solution onto the compact frequency range. 
\end{Corollary} \noindent
\Proof By making the interval larger and arguing for positive and negative frequencies separately,
it suffices to consider the case~$\omega_0=0$ and~$\omega_1>0$. Then, by choosing~$C$ sufficiently
close to one, we can arrange that~$\omega<\omega_{\max}$ with~$\omega_{\max}$ as in~\eqref{omegamaxcor}
with~$\varepsilon^2 = 1-C$. Then Corollary~\ref{corL12} gives the result.
\QED

We presented a first straightforward estimate of the series and showed that
it already allows us to derive interesting conclusions on the properties of solutions to the $1+1$-dimensional wave equation in the regime $E(\phi_-) \ll E(\phi)$.
%This simple estimate already captures all the essential features of the asymptotic behavior. 
In the following, we will demonstrate that the bound on the series~$g(\omega, \varepsilon)$ can be improved
substantially. The conclusion on the qualitative level, however, will remain the same. Therefore, these
improvements of the bounds are addressed more to technically-oriented readers.

\subsection{A First Improvement of the Estimate} \label{sec48}
In this section we give a first improvement of the estimate in Theorem~\ref{maintheoremsimple} by performing a more careful analysis of the series \eqref{eq:series}. These estimates are a preparation for the more advanced method for getting estimates which will be introduced in Section~\ref{secgoursat}.

For ease in notation we set
\beq \label{abdef}
a(\omega) = \frac{\log(4 \omega)}{2} \qquad \text{and} \qquad
b(\varepsilon) = 2\,|\log \varepsilon| \:.
\eeq
Then the series~\eqref{gseries} can be written as
\beq
g(a,b) := \sum_{n=0}^\infty \frac{1}{n!}\: \frac{1}{\sqrt{2n+1}}\:e^{(2n+3)\, a - \frac{b}{2n+3}} \:. \label{gabdef}
\eeq
Note that the last series converges absolutely and defines~$g$ as a smooth function on~$\R^2$.

Here is the main result of this section:
\begin{Thm} \label{thmmain2}
Let~$\phi$ be the solution of the Cauchy problem~\eqref{cauchy}. Assume that
\[ E(\phi_-^\eo) \leq \, \varepsilon^2 \: E(\phi^\eo) \:. \]
Then the initial data is small for small momenta in the sense that for all~$\omega \geq 0$,
\begin{align}
2\,\big| \hat{h}_\pm^\eo(\omega)\big|&\leq\big|\omega \hat{\phi}^\eo_0(\pm \omega) \big| +\; \big|\hat{\phi}^\eo_1(\pm \omega) \big| \notag \\
&\leq 12\, e^{4 \omega} \: \sqrt{E(\phi)} 
\:\max \bigg\{ \exp \Big( -\frac{1}{14}\: \frac{|\log \varepsilon|}{\sqrt{\omega}} \Big),\:
e\, \exp \Big(- \sqrt{|\log \varepsilon|} \:\Big) \bigg\} \:. \label{esprelim}
\end{align}
\end{Thm}
\Proof
In view of Proposition \ref{prop:series} and~\eqref{abdef}, \eqref{gabdef}, our task is to prove the following estimate,
\[ g(a,b) \leq 2\: e^{3a}\, \exp \big( e^{2a} \big) \:\max \bigg\{ \exp \Big( -\frac{b}{14}\: e^{-a} \Big),\:
\exp \bigg(1 - \sqrt{\frac{b}{2}} \:\bigg) \bigg\} \:. \]
We begin with the series~\eqref{gabdef}, leaving out the factor~$1/\sqrt{2n+1}$,
\[ g(a,b) \leq \sum_{n=0}^\infty \frac{1}{n!}\: e^{(2n+3)\,a - \frac{b}{2n+3}} \:. \]
We decompose this series into the sum over the first~$N$ summands and the remainder.
Estimating these two parts separately, we obtain
\begin{align*}
g(a,b) %\leq \sum_{n=0}^\infty &\frac{1}{n!}\: e^{(2n+3)\,a - \frac{b}{2n+3}}
&\leq \sum_{n=0}^N \frac{1}{n!}\: e^{(2n+3)\,a - \frac{b}{2n+3}}
+ \sum_{n=N+1}^\infty \frac{1}{n!}\: e^{(2n+3)\,a - \frac{b}{2n+3}} \\
&\leq e^{-\frac{b}{2N+3}} \sum_{n=0}^N \frac{1}{n!}\: e^{(2n+3)\,a}
+ \sum_{p=1}^\infty \frac{1}{(p+N)!}\: e^{(2p+2N+3)\,a - \frac{b}{2p+2N+3}} \\
&\leq e^{-\frac{b}{2N+3}} \: e^{3a}\, \exp \big( e^{2a} \big) 
+ e^{-\frac{b}{2N+3}} \:\frac{e^{(2N+3)\,a}}{N!}
\sum_{p=1}^\infty \frac{N!}{(p+N)!}\: e^{2pa - \frac{b}{2p+2N+3} + \frac{b}{2N+3}} \\
&\overset{(\ast)}{\leq} e^{-\frac{b}{2N+3}} \: e^{3a}\, \exp \big( e^{2a} \big) \bigg[ 1 + 
\sum_{p=1}^\infty \frac{N!}{(p+N)!}\: e^{2pa + \frac{2bp}{(2p+2N+3)(2N+3)}} \bigg] \\
&\leq e^{-\frac{b}{2N+3}} \: e^{3a}\, \exp \big( e^{2a} \big) \bigg[ 1 + 
\sum_{p=1}^\infty \bigg( \frac{1}{N+1}\: e^{2a + \frac{2b}{(2N+3)^2}} \bigg)^p \bigg] \:,
\end{align*}
where in~$(\ast)$ we used that
\[ \frac{e^{2Na}}{N!} \leq \sum_{n=0}^\infty \frac{e^{2na}}{n!} = \exp \big( e^{2a} \big) \:. \]
Choosing~$N$ so large that
\beq \label{Ncond}
\frac{1}{N+1}\: e^{2a + \frac{2b}{(2N+3)^2}} \leq \frac{1}{2}\:,
\eeq
we can compute the geometric series to obtain the estimate
\[ g(a,b) \leq 2\: 
e^{-\frac{b}{2N+3}} \: e^{3a}\, \exp \big( e^{2a} \big) \:. \]

In order to satisfy the condition~\eqref{Ncond}, we first choose
\[ \label{c1N}
2N+3 \geq \:\sqrt{2 b} \:, \]
which gives rise to the inequality
\[ e^{\frac{2b}{(2N+3)^2}} \leq e \:. \]
Moreover, choosing
\[ %\label{c2N}
N+1 \geq 2\: e^{2a + 1} \:, \]
we conclude that
\[ \frac{1}{N+1}\: e^{2a + \frac{b}{(2N+3)^2}} \leq \frac{1}{N+1}\: e^{2a + 1} \leq \frac{1}{2} \:, \]
implying that~\eqref{Ncond} holds. This leads us to choosing~$N$ as the integer in the range
\[ %\label{Nchoice}
N < \max \Big\{ 2\: e^{2a+1}, \sqrt{\frac{b}{2}}-\frac{1}{2} \Big\} \leq N+1 \:. \]
We thus obtain the estimates
\begin{align*}
2N+3 &\leq \max \Big\{ 4\: e^{2a+1}+3, \sqrt{2b}+2 \Big\} \\
g(a,b) &\leq 2\: e^{3a}\, \exp \big( e^{2a} \big) \:e^{-\frac{b}{2N+3}} \\
&\leq 2\: e^{3a}\, \exp \big( e^{2a} \big) \:\exp \bigg( -\frac{b}{\max \big\{ 4\: e^{a+1}+3, \sqrt{2b} +2 \big\} } \bigg) \\
&= 2\: e^{3a}\, \exp \big( e^{2a} \big) \:\max \bigg\{ \exp \Big( -\frac{b}{4\: e^{a+1}+3} \Big),\: 
\exp \Big( -\frac{b}{\sqrt{2b}+2} \Big) \bigg\} \:.
\end{align*}
Employing the inequalities
\[ \frac{1}{4\: e^{a+1}+3} \geq \frac{1}{14}\: e^{-a} \qquad \text{and} \qquad
\frac{b}{\sqrt{2b}+2} \geq \sqrt{\frac{b}{2}} -1 \]
gives the result.
\QED

We conclude this section with a comment on the parameter domains where the different estimates are better.
We first evaluate the point where the two arguments of the maximum coincide.
For simplicity disregarding the prefactor~$e$, we obtain
\[ \frac{1}{14}\: \frac{|\log \varepsilon|}{\sqrt{\omega}} =  \sqrt{|\log \varepsilon|} \quad
\Longleftrightarrow \quad \omega = \frac{|\log \varepsilon|}{196} \:. \]
We thus obtain the estimate
\[ \big| \hat{h}_\pm^\eo(\omega)\big|
\leq 24\,e\, e^{4 \omega} \: \sqrt{E(\phi)} \left\{
\begin{array}{cl} \displaystyle \exp \Big( -\frac{1}{14}\: \frac{|\log \varepsilon|}{\sqrt{\omega}} \Big) & \text{if } 
\displaystyle \omega > \frac{|\log \varepsilon|}{196} \\[1em]
\displaystyle \exp \Big(- \sqrt{|\log \varepsilon|} \:\Big) & \text{if } \displaystyle \omega \leq \frac{|\log \varepsilon|}{196} \:.
\end{array} \right. \]
For any given~$\omega$, one finds that~$|\hat{h}_\pm^\eo(\omega)| \lesssim \exp(-\sqrt{|\log \varepsilon|})$
asymptotically as~$\varepsilon \searrow 0$. This is a faster decay than the
asymptotics~$|\hat{h}_\pm^\eo(\omega)| \lesssim 1/\sqrt{|\log \varepsilon|}$ as obtained in Theorem~\ref{maintheoremsimple}. On the other hand, fixing~$\varepsilon$
and considering the asymptotics~$\omega \rightarrow \infty$, the estimate of
Theorem~\ref{maintheoremsimple} is slightly better than that of Theorem \ref{thmmain2}
because of the factor~$|\log \varepsilon|^{-\frac{1}{2}}$ in~\eqref{initestimate2}.
However, in this limiting regime, both theorems are not useful, because the
estimates are worse than the simple pointwise bound of Lemma~\ref{constantbound}.
With this in mind, the above theorems are useful only for~$\omega$ 
in a finite interval and for small~$\varepsilon$.

We now turn to substantially more sophisticated techniques to obtain the best estimate in this paper
(see Corollary~\ref{cormain}).

\subsection{Formulation as a Goursat Problem for the Klein-Gordon Equation} \label{secgoursat}
We now develop another method for estimating the series~$g$ in~\eqref{gseries}.
This method is based on the observation that~$g$ is a solution of a
partial differential equation in~$\varepsilon$ and~$\omega$.
As we shall see, this PDE is indeed the Klein-Gordon equation (see~\eqref{KG} below),
and the above series is obtained as the solution of a characteristic initial value problem
(usually referred to as Goursat problem; see Proposition~\ref{prpgoursat} below).
This observation makes it possible to analyze the series in~\eqref{gseries} with familiar methods
of hyperbolic PDEs, as will be worked out in Sections~\ref{seccontour}--\ref{secescont}.
Before entering the constructions, we remark that there seems no direct relation between
the original wave equation and the PDE in~$\varepsilon$ and~$\omega$.
To our knowledge, it is not even clear why~$g$ satisfies a PDE, and why
this PDE is hyperbolic.

We again work with the parameters~$a$ and~$b$ as introduced in~\eqref{abdef}.
Differentiating the function~$g(a,b)$ in~\eqref{gabdef} with respect to~$a$ and~$b$ gives
\begin{align*}
\partial_a g(a,b) &= \sum_{n=0}^\infty \frac{1}{n!}\: \frac{1}{\sqrt{2n+1}}\:(2n+3)\: e^{(2n+3)\, a - \frac{b}{2n+3}} \\
\partial_b \partial_a g(a,b) &= \sum_{n=0}^\infty \frac{1}{n!}\: \frac{1}{\sqrt{2n+1}}\:
\Big(-\frac{2n+3}{2n+3} \Big)\: e^{(2n+3)\, a - \frac{b}{2n+3}} = -g(a,b)\:.
\end{align*}
Hence~$g$ is a solution of the PDE
\beq \label{KG}
(\partial_a \partial_b + 1)\, g = 0 \:.
\eeq
This is the $(1+1)$-dimensional Klein-Gordon equation of mass one in light cone coordinates.
Introducing the coordinates
\begin{align*}
T &= a+b \:,& X&=a-b \\
\partial_T &= \frac{1}{2} \big( \partial_a + \partial_b \big) \:,& 
\partial_X &= \frac{1}{2} \big( \partial_a - \partial_b \big) \:,
\end{align*}
the equation takes the more familiar form
\[ \big(\partial_T^2 - \partial_X^2 + 1 \big) \,g = 0 \:. \]
This PDE comes with initial conditions at~$b=0$ given by the series
\beq \label{g0def}
g_0(a) := g(a,0) = \sum_{n=0}^\infty \frac{1}{n!}\: \frac{1}{\sqrt{2n+1}}\:e^{(2n+3)\, a} \:.
\eeq
Moreover, Lebesgue's monotone convergence theorem implies that
\beq \label{decay}
\lim_{b \rightarrow \infty} g(a,b) = \lim_{a \rightarrow -\infty} g(a,b) = 0 \:.
\eeq
The above PDE together with the initial conditions determine the function~$g$ uniquely:

\begin{Prp} \label{prpgoursat} The Goursat problem
\beq \label{goursat}
\big( \partial_a \partial_b + 1 \big)\,g(a,b) = 0 \:,\qquad g(a,0) = g_0(a)
\eeq
together with the decay conditions~\eqref{decay} has a unique solution in the half space
\[ (a,b) \in \R \times \R^+_0 \:. \]
It has the integral representation
\beq \label{gabform}
g(a,b) = \int_{-\infty}^a J_0\Big( 2\,\sqrt{(a-\tau) \,b} \,\Big)\: g_0'(\tau)\: d\tau \:.
\eeq
\end{Prp}
\Proof The appearance of the Bessel function in~\eqref{gabform} can be
understood directly from the form of the Green's kernels of the Klein-Gordon
equation as given in~\eqref{Swedge} and~\eqref{Svee}.
Indeed, choosing the space-time coordinates~$(T, X)$ and setting the mass to one,
the causal fundamental solution~\eqref{Kdef} takes the form
\[ %\label{KTX}
K_1(T,X) =  -\frac{i}{4 \pi} \: \epsilon(T) \:\Theta\big(T^2-X^2 \big) \:
J_0 \Big( \sqrt{T^2-X^2} \Big) \]
(where~$\epsilon$ is again the sign function).
Hence in light-cone coordinates,
\beq \label{Kform}
K_1(T,X) = K[a,b] := -\frac{i}{4 \pi} \: \Theta(ab)\: \epsilon(b)\: J_0\Big( 2\, \sqrt{ab} \Big)
\eeq
(note that~$T^2-X^2=(a+b)^2 - (a-b)^2 = 4 ab$).
It is a solution of the homogeneous Klein-Gordon equation. Hence also the convolution integral
\[ h(a,b) := 4 \pi i \int_{-\infty}^\infty K[a-\tau, b]\: g_0'(\tau)\: d\tau \]
satisfies the Klein-Gordon equation. Using the explicit form of~$K_1$ in~\eqref{Kform}, one
sees that the function~$h$ coincides with the function~$g$ in~\eqref{gabform}.

Let us verify that the function~$h$ has the desired boundary values at~$b=0$.
Using that~$J_0(0)=1$, we obtain
\begin{align*}
\lim_{b \searrow 0} h(a,b) &= \lim_{b \searrow 0} \int_{-\infty}^a J_0\Big( 2\, \sqrt{(a-\tau)\,b}\, \Big)\: g_0'(\tau)\: d\tau \\
&= \int_{-\infty}^a g_0'(\tau)\: d\tau = g_0(a)\:,
\end{align*}
where we made use of the fact that~$g_0(\tau)$ vanishes as~$\tau \rightarrow -\infty$.

It remains to show uniqueness. Let~$\tilde{g}$ be another solution of the Klein-Gordon equation
with the same boundary values at~$b=0$. Then the difference~$\phi:=g-\tilde{g}$ is a solution which vanishes
at~$b=0$. Our task is to prove that $\phi$ vanishes identically.
This result can be understood intuitively from the fact that, being massive, a Klein-Gordon wave propagates
with subluminal speed, implying that if it were non-zero, it would intersect the null line~$b=0$.
In order to prove this result, we consider the Fourier representation of~$\phi$,
\[ \phi(T,X) = \int_{-\infty}^\infty \Big( \hat{\phi}_+(k)\: e^{-i \omega(k)\,T} + \hat{\phi}_-(k)\: e^{i \omega(k)\,T} \Big)
\:e^{ikX} \:, \]
where~$\omega(k):= \sqrt{k^1+1}$. The fact that~$\phi$ vanishes on the line~$b=0$ implies that
\[ 0 = \phi(a,a) = \int_{-\infty}^\infty \Big( \hat{\phi}_+(k)\: e^{-i \omega(k)\,a} + \hat{\phi}_-(k)\: e^{i \omega(k)\,a} \Big)
\:e^{ika} \:. \]
Multiplying by~$e^{i p a}$ and integrating over~$a$, we obtain zero for any value of~$p$.
Since the mappings
\[ \R \mapsto \R^\pm \:,\qquad k \mapsto k \pm \omega(k) \]
are both injective, it follows that the functions~$\hat{\phi}_\pm$ are both zero.
Hence~$\phi$ vanishes identically.
\QED

We remark that the identity~\eqref{gabform} can also be derived without referring to hyperbolic PDEs
simply by manipulating the power series; for details see Appendix~\ref{appA}.

\subsection{Arranging Initial Data in Closed Form} \label{secesinit}
The initial data as given by the series~\eqref{g0def} has the disadvantage that
it is not a simple explicit function. In view of the fact that the integral representation~\eqref{gabform}
involves the derivative of~$g_0$ and that the Bessel function has an oscillatory behavior, it is not obvious
how an estimate of the initial data translates into a corresponding estimate of the solution.
For this reason, it is preferable to estimate the solution in terms of new solutions
of the Goursat problem~\eqref{goursat} for initial data given in closed form.

\begin{Lemma} \label{lemmaschwarz}
The solution of the Goursat problem~\eqref{goursat} with initial data~\eqref{g0def}
satisfies the inequality
\[ \big| g(a,b) \big| \leq \sqrt{ g^{(1)}(a,b) \, g^{(2)}(a,b) } \:, \]
where the functions~$g^{(1)}$ and~$g^{(2)}$ are solutions of the Goursat problem~\eqref{goursat}
corresponding to the initial data
\beq
g_0^{(1)}(a) = e^{3a} \,\exp \big( e^{2a} \big) \qquad \text{and} \qquad
g_0^{(2)}(a) = e^{3a} \int_0^1 \exp \big( s^2 \,e^{2a} \big)\: ds \:, \label{g012}
\eeq
respectively.
\end{Lemma}
\Proof Since all summands in the series~\eqref{gabdef} are non-negative, the Schwarz inequality gives
\begin{align*}
g(a,b) &= \sum_{n=0}^\infty \bigg( \frac{1}{n!}\: e^{(2n+3)\, a - \frac{b}{2n+3}} \bigg)^\frac{1}{2} 
\; \bigg( \frac{1}{n!}\: \frac{1}{2n+1}\:e^{(2n+3)\, a - \frac{b}{2n+3}} \bigg)^\frac{1}{2} \\
&\leq \bigg( \sum_{n=0}^\infty \frac{1}{n!}\: e^{(2n+3)\, a - \frac{b}{2n+3}} \bigg)^\frac{1}{2}
\bigg( \sum_{n=0}^\infty \frac{1}{n!}\: \frac{1}{2n+1} \:e^{(2n+3)\, a - \frac{b}{2n+3}} \bigg)^\frac{1}{2} \:.
\end{align*}
By direct inspection one sees that 
each bracket is a solution of the Goursat problem~\eqref{goursat} corresponding to the initial data
\begin{align*}
g_0^{(1)}(a) &= \sum_{n=0}^\infty \frac{1}{n!}\: e^{(2n+3)\, a}
= e^{3a} \:\sum_{n=0}^\infty \frac{1}{n!}\: \big( e^{2a} \big)^n
= e^{3a} \,\exp \big( e^{2a} \big) \qquad \text{and} \\
g_0^{(2)}(a) &= \sum_{n=0}^\infty \frac{1}{n!}\: \frac{1}{2n+1} \:e^{(2n+3)\, a}
= e^{3a} \:\sum_{n=0}^\infty \frac{1}{n!}\: \frac{1}{2n+1} \:\big( e^a \big)^{2n} \\
&= e^{3a} \int_0^1 \bigg( \sum_{n=0}^\infty \frac{1}{n!}\: s^{2n} \:\big( e^a \big)^{2n}  \bigg)\: ds
= e^{3a} \int_0^1 \exp \big( s^2 \,e^{2a} \big)\: ds \:,
\end{align*}
respectively. This concludes the proof.
\QED

\subsection{Reformulation as a Contour Integral} \label{seccontour}
In this section, we rewrite the integral representation~\eqref{gabform} 
in Proposition~\ref{prpgoursat} as a contour integral.
We make use of the fact that the Bessel function
in~\eqref{gabform} also arises in the causal fundamental solution~\eqref{Kform},
which in turn can be represented in momentum space by a distribution supported on the
mass shell. Our starting point is the formula~\eqref{gabform}.
Introducing the integration variable
\[ q := 2\,\sqrt{(a-\tau) \,b} \:, \]
we obtain
\[ a-\tau = \frac{q^2}{4b} \:,\qquad d\tau = \frac{1}{2b}\: q\: dq \]
and thus
\[ g(a,b) = \frac{1}{2b} \int_0^\infty J_0(q)\: g_0'\Big( a- \frac{q^2}{4b} \Big)\: q\: dq \:. \]
Since both functions~$J_0$ and~$g_0'$ are even in~$t$, we can write this integral as
\beq \label{gabq}
g(a,b) = \frac{1}{4b} \int_{-\infty}^\infty \Big( J_0(q)\: \epsilon(q) \Big) \: 
\Big( g_0'\Big( a- \frac{q^2}{4b} \Big)\: q \Big)\: dq \:.
\eeq

Using Plancherel's theorem, we can also compute this inner product in momentum space.
In preparation, we compute the Fourier transform of the Bessel function:
\begin{Lemma} \label{lemmaJhat} For any~$p \in \R$,
\begin{align*}
\int_{-\infty}^\infty J_0(q)\: \epsilon(q)\: e^{i p q}\: dq
= 2 i \:\frac{\epsilon(p)}{\sqrt{p^2-1}}\: \chi_{\R \setminus [-1,1]}(p)
\end{align*}
(where~$\chi$ denotes the characteristic function, and~$\epsilon$ is again the sign function).
\end{Lemma}
\Proof According to~\eqref{Kform} and~\eqref{Kfourier}, for any~$q \in \R$,
\begin{align*}
J_0(q)\: \epsilon(q) &= 4 \pi i\, K_1 \big(T=q,X=0\big) = 4 \pi i \int \frac{d\omega\, dk}{(2 \pi)^2}\:
\delta \big( \omega^2 - k^2-1\big)\, \epsilon(\omega)\: e^{-i \omega q} \\
&= \frac{i}{\pi} \int_{-\infty}^\infty d\omega \: \epsilon(\omega)\:
e^{-i \omega q} \int_{-\infty}^\infty \delta \big(\omega^2-k^2-1 \big)\: dk \\
&= \frac{i}{\pi} \int_{\R \setminus [-1,1]} \frac{\epsilon(\omega)}{\sqrt{\omega^2-1}}\: e^{-i \omega q}\: d\omega \:.
\end{align*}
We now apply Plancherel's theorem.
\QED

\begin{Prp} \label{prpgab}
The function~$g(a,b)$ in~\eqref{gabq} can be written as
\beq \label{gint}
g(a,b) = \frac{1}{\pi} \int_{\sqrt{2b}}^\infty \frac{k}{\sqrt{k^2-2b}}\:  \hat{g}(a, k) \: dk
\eeq
with
\beq \label{hgdef}
\hat{g}(a,k) := \int_{-\infty}^\infty  g_0 \Big( a- \frac{y^2}{2} \Big)\: e^{i k y}\: dy \:.
\eeq
\end{Prp}
\Proof Applying Plancherel's theorem to~\eqref{gabq} gives
\beq \label{gabfourier}
g(a,b) = \frac{1}{4b} \int_{-\infty}^\infty \frac{dp}{2 \pi}\: \hat{J}(-p) \: \hat{h}_\pm(p) \:,
\eeq
where
\begin{align*}
\hat{J}(p) &:= \int_{-\infty}^\infty J_0(q)\: \epsilon(q)\: e^{i p q}\: dq \\
\hat{h}_\pm(p) &:= \int_{-\infty}^\infty g_0'\Big( a- \frac{q^2}{4b} \Big)\: q\: e^{i p q}\: dq
\end{align*}
(this relation is verified most easily by substituting the last two equations into~\eqref{gabfourier}
and using that~$\int_{-\infty}^\infty e^{i p r} dp = 2 \pi \delta(r)$).
The first Fourier integral was computed in Lemma~\ref{lemmaJhat}.
The second Fourier integral can be simplified using integration by parts,
\[ \hat{h}_\pm(p)
= -2b \int_{-\infty}^\infty  e^{i p q}\:\frac{d}{dq} g_0\Big( a- \frac{q^2}{4b} \Big)\: dq
= ip\:2b \int_{-\infty}^\infty  g_0\Big( a- \frac{q^2}{4b} \Big)\: e^{i p q}\: dq \:. \]
Introducing the new integration variable~$y=q/\sqrt{2b}$ gives
\begin{align*}
\hat{h}_\pm(p) 
&= \sqrt{8}\: ip\, b^{\frac{3}{2}} \int_{-\infty}^\infty  g_0 \Big( a- \frac{y^2}{2} \Big)\: e^{i \tilde{p} y}\: dy 
\qquad \text{with} \qquad \tilde{p} := \sqrt{2b}\: p \\
&= \sqrt{8}\: ip\, b^{\frac{3}{2}} \: \hat{g}\big(a, \sqrt{2b}\, p \big) \:,
\end{align*}
where in the last step we used the notation~\eqref{hgdef}.

Combining the above formulas, we obtain
\begin{align*}
g(a,b) &= \frac{1}{4b} \int_{\R \setminus [-1,1]} \frac{dp}{2 \pi}\: (-2 i)\,\frac{\epsilon(p)}{\sqrt{p^2-1}}\: 
\sqrt{8}\, ip\, b^{\frac{3}{2}} \: \hat{g}\big(a, \sqrt{2b}\, p \big) \\
&= \frac{\sqrt{2b}}{2\pi} \int_{\R \setminus [-1,1]} \frac{|p|}{\sqrt{p^2-1}}\: 
\hat{g}\big(a, \sqrt{2b}\, p \big) \: dp \\
&= \frac{\sqrt{2b}}{\pi} \int_1^\infty \frac{p}{\sqrt{p^2-1}}\: 
\hat{g}\big(a, \sqrt{2b}\, p \big) \: dp 
= \frac{1}{\pi} \int_{\sqrt{2b}}^\infty \frac{k}{\sqrt{k^2-2b}}\: 
\hat{g}(a, k) \: dk  \:,
\end{align*}
where in the last line we used that the integrand is even.
\QED

\subsection{Estimates of the Contour Integral} \label{secescont}
Our next goal is to estimate the contour integral in~\eqref{hgdef}.
In view of the estimate of Lemma~\ref{lemmaschwarz}, for the function~$g_0$
it suffices to consider the explicit functions~$g_0^{(1)}$ and~$g_0^{(2)}$ in~\eqref{g012}.
In order to treat these two functions together, for a given parameter~$s \in [0,1]$ we choose
\beq \label{g0combine}
g_0(a) = e^{3a} \:\exp \big( s^2 \,e^{2a} \big) \:.
\eeq
Clearly, setting~$s=1$ gives the function~$g_0^{(1)}$.
In order to treat the function~$g_0^{(2)}$, we will later integrate over the parameter~$s \in [0,1]$
(see Section~\ref{secsintegral}). Thus we turn our attention to estimating the integral
\[ \hat{g}(a,k) = \int_{-\infty}^\infty  g_0 \Big( a- \frac{y^2}{2} \Big)\: e^{i k y}\: dy \]
for the function~$g_0$ as given by~\eqref{g0combine}.
In order to simplify the notation, we set
\beq \label{lambdadef}
\lambda = s^2 \,e^{2a} \:.
\eeq
Then the transformation
\[ \exp \Big( s^2\, e^{2 \big(a-\frac{y^2}{2} \big)} \Big) = \exp \big( \lambda\, e^{-y^2} \big) \]
allows us to rewrite the above integral as
\beq \label{ghdef}
\hat{g}(a,k) = e^{3a} \int_{-\infty}^\infty \exp \Big(-\frac{3}{2}\: y^2 + \lambda \,e^{-y^2} + i k y \Big) \: dy \:.
\eeq
We also write this integral as
\begin{align}
\hat{g}(a,k) &= e^{3a} \int_{-\infty}^\infty e^{\chi(y)}\: dy \qquad \text{with} \label{Idef} \\
\chi(y) \:&\!:= -\frac{3}{2}\: y^2 + \lambda \,e^{-y^2} + i k y \:.
\end{align}

We want to apply a saddle-point argument. To this end, we first compute the critical points of the function~$\chi$.
In fact, a straightforward computation shows that there is only one critical point, which lies on the
imaginary axis at
\[ y = i \beta \:, \]
where~$\beta$ is defined implicitly by the equation
\beq \label{krep}
k = 3\, \beta + 2 \lambda\,\beta\: e^{\beta^2} \:.
\eeq
Our strategy is to deform the integration contour such that it goes through this critical point.
For simplicity, we choose the integration contour as a straight line parallel to the real axis,
\[ %\label{gammadef}
y = \gamma(t) := t + i \beta \:. \]
We thus obtain
\begin{align*}
\chi(y) &= \lambda \, e^{-t^2 + \beta^2- 2 i \beta t} - 2 \lambda\,e^{\beta^2}\:\beta^2 
-\frac{3}{2}\: \beta^2 - \frac{3}{2}\: t^2 \\
&\quad\: + 2 i\,\lambda \,e^{\beta^2}\:\beta\: t \:,
\end{align*}
and thus
\[ e^{\chi(y)} = A\, \exp \bigg\{ C\, e^{-t^2}\, e^{-2 i \beta t} \bigg\} \:B(t) \:, \]
where used~\eqref{krep} in order to express~$k$ in terms of~$\beta$ and set
\begin{align}
A &= \exp \Big( - 2 \lambda \,e^{\beta^2}\:\beta^2 -\frac{3}{2}\: \beta^2 \Big) \label{Adef} \\
B(t) &= e^{2 i\,\lambda \,e^{\beta^2}\:\beta\: t}\: \exp \Big(- \frac{3}{2}\: t^2 \Big) \label{Bdef} \\
C &= \lambda \,e^{\beta^2} \label{Cdef} \:.
\end{align}
Using this formula in~\eqref{Idef}, we can decompose the integral as
\begin{align}
\hat{g}(a,k)  &= e^{3a}\: A \,\mathscr{J} \qquad \text{with} \label{AIJ} \\
\mathscr{J} \,&\!:= \int_{-\infty}^\infty \exp \Big\{ C\, e^{-t^2}\, e^{-2i \beta t}  \Big\}\: B(t)\: dt \:. \label{I2}
\end{align}
In order to estimate this integral, we first take the absolute value of the integrand
\begin{align}
|\mathscr{J}| &\leq \int_{-\infty}^\infty \bigg| \exp \Big\{ C\, e^{-t^2}\, e^{-2i \beta t}  \Big\} \bigg|\;
e^{-\frac{3}{2}\, t^2}\: dt \notag \\
&= \int_{-\infty}^\infty \exp \Big\{ C\, e^{-t^2}\, \re e^{-2i \beta t}  \Big\} \;
e^{-\frac{3}{2}\, t^2}\: dt
\leq  \int_{-\infty}^\infty \exp \Big\{ C\, e^{-t^2} \Big\} \;
e^{-\frac{3}{2}\, t^2}\: dt \:. \label{Jes}
\end{align}
The obtained integral is estimated further in the next lemma.
\begin{Lemma} For any~$C \geq 0$,
\beq \label{intest}
\int_0^\infty \exp \Big\{ C\, e^{-t^2} \Big\}\: e^{-\frac{3}{2}\, t^2}\: dt
\leq 2\: \frac{e^C}{\sqrt{1+C}} \:.
\eeq
\end{Lemma}
\Proof For~$t \in [0,1]$ we estimate the inner exponential by a polynomial,
\[ e^{-t^2} \leq 1 - t^2 + \frac{t^4}{2}\: \sup_{\xi \in [0,1]} e^{-\xi^2} 
\leq 1 - t^2 + \frac{t^4}{2} \leq 1 - \frac{t^2}{2} \:. \]
This gives the estimate
\beq
\int_0^1 \exp \Big\{ C\, e^{-t^2} \Big\}\: e^{-\frac{3}{2}\, t^2}\: dt 
\leq \int_0^1 e^C \,\exp \Big\{ -\frac{C}{2}\:t^2 \Big\} \: e^{-\frac{3}{2}\, t^2}\: dt
\leq \sqrt{\frac{\pi}{2}}\: \frac{e^C}{\sqrt{C+3}} \:. \label{first}
\eeq

In the remaining parameter range~$t \in [1,\infty)$, we use that~$e^{-t^2} < e^{-1}$ to obtain
\[ \int_1^\infty \exp \Big\{ C\, e^{-t^2} \Big\}\: e^{-\frac{3}{2}\, t^2}\: dt 
\leq e^{\frac{C}{e}} \int_0^\infty e^{-\frac{3}{2}\, t^2}\: dt
= \sqrt{\frac{\pi}{6}}\: e^{\frac{C}{e}} \:. %\label{second}
\]

For large values of~$C$, the contribution~\eqref{first} clearly dominates. Since this
contribution has no zeros and all contributions are bounded near~$C=0$, one finds that~\eqref{intest}
holds with some numerical constant on the right side. By direct inspection one sees that this constant can be chosen
equal to two.
\QED
Combining the above estimates, we obtain the following result.

\begin{Lemma} \label{lemmages} The integral~\eqref{ghdef} can be estimated by
\begin{align}
\big| \hat{g}(a,k) \big| &\leq \frac{c\:e^{3a}}{\sqrt{1 + \lambda \, e^{\beta^2}}}\:
e^{-h(\lambda,k)} \qquad \text{with} \label{ghes} \\
h(\lambda, k) \:&\!:= \frac{3}{2}\: \beta^2 - \lambda \, e^{\beta^2} 
\,\Big( 1 - 2\, \beta^2 \Big) \:, \label{hdef}
\end{align}
where~$c$ is a numerical constant, $\lambda$ is defined by~\eqref{lambdadef},
and~$\beta$ is given implicitly by~\eqref{krep}.
\end{Lemma}
\Proof We combine~\eqref{Jes} with~\eqref{intest} 
and apply the resulting inequality in~\eqref{AIJ}.
Using~\eqref{Cdef} gives the result.
\QED

We finally collect a few properties of the function~$h$
in~\eqref{hdef} which will be needed in the next section.
\begin{Lemma} For any fixed~$\lambda$,
\begin{align}
h(\lambda, k) &= -\frac{3}{2} \beta^2 - k \:\Big( \frac{1}{2\beta} - \beta \Big) + \frac{3}{2} \label{halt} \\
\frac{\partial h(\lambda, k)}{\partial k} &= \beta  \label{dhdk} \\
\frac{\partial h(\lambda, k)}{\partial \lambda} &= -e^{\beta^2} \label{dhdl} \\
\frac{\partial^2 h(\lambda, k)}{\partial \lambda^2} &>0 \label{d2hdl2} \:,
\end{align}
where $k$ is given via~\eqref{krep} in terms of~$\lambda$ and~$\beta$. Moreover,
for any~$k>\tilde{k}$,
\beq \label{hes}
h(\lambda, k) \geq h(\lambda, \tilde{k}) + \tilde{\beta}\: \big(k-\tilde{k} \big)\:.
\eeq
\end{Lemma}
\Proof The relation~\eqref{halt} follows immediately from~\eqref{hdef} and~\eqref{krep}.
Next, a direct computation using again~\eqref{hdef} and~\eqref{krep} yields
\begin{align}
\frac{\partial h}{\partial \beta} &= 3\, \beta + 2 \lambda\: \beta\: e^{\beta^2}
+ 4 \lambda\: \beta^3\: e^{\beta^2} \\
\frac{\partial k}{\partial \beta} &= 3 + 2 \lambda\: e^{\beta^2}
+ 4 \lambda\: \beta^2\: e^{\beta^2} \:. \label{dkdimy0}
\end{align}
Combining these equations with the chain rule gives~\eqref{dhdk}.

In order to compute the partial derivatives with respect to~$\lambda$, we first compute the
total derivative of~\eqref{krep} for fixed~$k$,
\[ 0 = dk = 2 \beta\, e^{\beta^2} \: d\lambda + \Big( 3 + 2 \lambda e^{\beta^2} (1+2 \beta^2) \Big)\: d\beta \:. \]
Hence
\beq \label{D1}
\frac{d \beta}{d\lambda} = -\frac{2 \beta\, e^{\beta^2}}{3 + 2 \lambda e^{\beta^2} (1+2 \beta^2)} \:.
\eeq
This formula shows in particular that, for fixed~$k$, the function~$\beta$ is monotone decreasing in~$\lambda$.
On the other hand, a direct computation using~\eqref{halt} and again~\eqref{krep} gives
\beq \label{D2}
\frac{\partial h}{\partial\beta}
= \frac{3 + 2 \lambda e^{\beta^2} (1+2 \beta^2)}{2\, \beta}
\eeq
(the partial derivative is again computed for fixed~$k$).
Taking the product of~\eqref{D1} and~\eqref{D2} gives~\eqref{dhdl}.
Differentiating once again and using that~$\beta$ is monotone
decreasing gives~\eqref{d2hdl2}.

In order to derive~\eqref{hes}, we first note that from~\eqref{krep} or~\eqref{dkdimy0}
it follows that, for fixed~$\lambda$, the function~$\beta$ is monotone increasing in~$k$. Therefore,
\[ h(\lambda, k) - h(\lambda, \tilde{k}) = \int_{\tilde{k}} ^k
\frac{\partial h(\lambda, \hat{k})}{\partial \hat{k}}\: d\hat{k}
\overset{\eqref{dhdk}}{=} \int_{\tilde{k}}^k
\hat{\beta}\: d\hat{k} \geq \tilde{\beta}\: \big(k-\tilde{k} \big)\:. \]
This concludes the proof.
\QED

\subsection{Estimate of~\texorpdfstring{$g^{(1)}$}{gone}} \label{seckintegral} 
The goal of this section is to estimate the solution of the Goursat problem~$g(a,b)$
in~\eqref{goursat} with initial data~$g_0^{(1)}$ as in~\eqref{g012}.
Our starting point is the estimate of Lemma~\ref{lemmages}, where we set~$s=1$
(cf.~\eqref{g0combine} and~\eqref{g012}). Our task is to estimate
the integral~\eqref{gint}. To this end, we need to distinguish different cases: \\

\noindent
{\bf{Case~(A)}}: $0 \leq \beta < 1$. In view of~\eqref{krep}, this corresponds to the range for~$k$
\beq \label{kloges}
k < k_0 := 3 + 2 e\, \lambda \:.
\eeq
In this case, we can estimate~$\beta$ in terms of~$k$ by
\beq \label{imy0A}
k \leq (3 + 2 e \lambda)\, \beta \:,\qquad
\beta \geq \frac{k}{3 + 2 e \lambda} \:.
\eeq

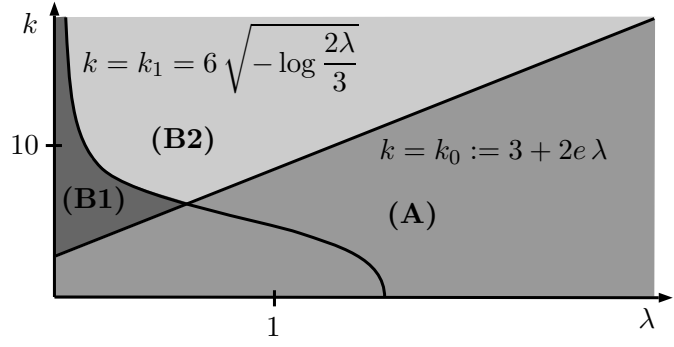
\begin{figure}
% \usepackage[usenames,dvipsnames]{pstricks}
% \usepackage{epsfig}
% \usepackage{pst-grad} % For gradients
% \usepackage{pst-plot} % For axes
% \usepackage[space]{grffile} % For spaces in paths
% \usepackage{etoolbox} % For spaces in paths
% \makeatletter % For spaces in paths
% \patchcmd\Gread@eps{\@inputcheck#1 }{\@inputcheck"#1"\relax}{}{}
% \makeatother
% 
\psscalebox{1.0 1.0} % Change this value to rescale the drawing.
{
\begin{pspicture}(-2.5,-2.213948)(16.016666,2.213948)
\definecolor{colour0}{rgb}{0.8,0.8,0.8}
\definecolor{colour1}{rgb}{0.6,0.6,0.6}
\definecolor{colour2}{rgb}{0.4,0.4,0.4}
\pspolygon[linecolor=colour0, linewidth=0.02, fillstyle=solid,fillcolor=colour0](0.7266667,2.0919895)(8.526667,2.0969896)(2.3366666,-0.3680104)(1.8666667,-0.2230104)(1.5116667,-0.0930104)(1.2916666,0.036989592)(1.0566666,0.27198958)(0.9066667,0.5819896)(0.81166667,1.0169896)(0.76666665,1.5969896)
\pspolygon[linecolor=colour1, linewidth=0.02, fillstyle=solid,fillcolor=colour1](0.6016667,-1.6330104)(0.6116667,-1.1030104)(8.5616665,2.0619895)(8.551666,-1.6080104)
\pspolygon[linecolor=colour2, linewidth=0.02, fillstyle=solid,fillcolor=colour2](0.58166665,2.0869896)(0.59166664,-1.0630105)(2.3616667,-0.3780104)(1.9166666,-0.2430104)(1.4416667,-0.083010405)(1.0716667,0.20698959)(0.90166664,0.5969896)(0.7916667,1.1169896)(0.76166666,1.6269896)(0.7316667,2.0969896)
\rput[bl](5.0066667,-0.7380104){\normalsize{\bf{(A)}}}
\psline[linecolor=black, linewidth=0.04, arrowsize=0.05291667cm 2.0,arrowlength=1.4,arrowinset=0.0]{->}(0.58666664,-1.6230104)(0.58666664,2.3069897)
\psline[linecolor=black, linewidth=0.04, arrowsize=0.05291667cm 2.0,arrowlength=1.4,arrowinset=0.0]{->}(0.58666664,-1.6230104)(8.806666,-1.6330104)
\rput[bl](0.15833333,1.8886563){\normalsize{$k$}}
\rput[bl](8.351666,-2.0313437){\normalsize{$\lambda$}}
\rput[bl](0.67,-0.5430104){\normalsize{\bf{(B1)}}}
\rput[bl](1.8566667,0.2569896){\normalsize{\bf{(B2)}}}
\rput[bl](4.9066668,0.1519896){\normalsize{$k=k_0 := 3 + 2e\, \lambda$}}
\rput[bl](0.95666665,1.080323){\normalsize{$\displaystyle k = k_1 = 6\, \sqrt{-\log \frac{2 \lambda}{3} }$}}
\psbezier[linecolor=black, linewidth=0.04](0.7366667,2.0969896)(0.78302014,1.1752687)(0.80407864,0.49982804)(1.2066667,0.09698959350585938)(1.6092547,-0.30584887)(3.37802,-0.6147313)(3.8466666,-0.77301043)(4.3153133,-0.9312895)(4.9792547,-1.0958489)(4.9766665,-1.6330104)
\psline[linecolor=black, linewidth=0.04](0.6016667,-1.0780104)(8.5616665,2.0819895)
\psline[linecolor=black, linewidth=0.04](3.5116668,-1.5330104)(3.5116668,-1.7480104)
\psline[linecolor=black, linewidth=0.04](0.47166666,0.3919896)(0.68666667,0.3919896)
\rput[bl](3.4,-2.1380105){\normalsize{$1$}}
\rput[bl](0.0,0.2819896){\normalsize{$10$}}
\end{pspicture}
}
\caption{Different cases in the~$k\lambda$-plane.}
\label{figcases}
\end{figure}

\noindent
{\bf{Case~(B)}}: $\beta \geq 1$. In view of~\eqref{krep}, this corresponds to the range for~$k$
\[ k \geq k_0 = 3 + 2 e\, \lambda \:. \]
In order to express~$\beta$ in terms of~$k$, we distinguish two sub-cases.
We set
\beq \label{y1def}
\im y_1 := \left\{ \begin{array}{cl} 
\displaystyle \sqrt{-\log \frac{2 \lambda}{3}} & \text{if~$\displaystyle \lambda<\frac{3}{2 e}$} \\[0.8em]
1 & \text{if~$\displaystyle \lambda \geq \frac{3}{2 e} $\:.} \end{array} \right.
\eeq
\begin{itemize}[leftmargin=2em]
\item[] \hspace*{-2em} {\bf{Case~(B1)}}: $1 \leq \beta < \im y_1$.
Clearly, this case only occurs if~$\im y_1 >1$, which by~\eqref{y1def} implies that
\[ \lambda < \frac{3}{2e}\:. \]
Moreover,
\[ %\label{lB1}
\lambda\: e^{\beta^2} \leq \lambda\: e^{\im^2 y_1} = \frac{3}{2} \:. \]
Using~\eqref{krep}, we obtain
\begin{align*}
k < k_1 \,\,&\!\!:= 3\, \im y_1 + 2 \lambda\,\im y_1\: e^{\im^2 y_1} \\
&= \sqrt{-\log (2\lambda/3)} \:\big( 3 + 2 \lambda\, e^{-\log (2\lambda/3)} \big) = 6\, \sqrt{-\log (2 \lambda/3)} 
= 6\, \im y_1\:.
\end{align*}
Therefore, we can estimate~\eqref{krep} from above and below by
\begin{gather}
k - 3 \beta \leq 3\, \beta \label{kB1} \\
3 \beta \leq k \leq 6\, \beta \:,\qquad
\frac{k}{6} \leq \beta \leq \frac{k}{3}\:. \label{imy0B1}
\end{gather}
\item[] \hspace*{-2em} {\bf{Case~(B2)}}: $\beta \geq \max\{1,\im y_1\}$. In this case,
\[ %\label{lB2}
\lambda\: e^{\beta^2} \geq \lambda\: e^{\im^2 y_1} = \frac{3}{2} \:, \]
making it possible to estimate~\eqref{krep} by
\begin{gather}
k - 3 \beta \geq 3\, \beta \label{kB2} \\
k = 3\, \beta + 2 \lambda\,\beta\: e^{\beta^2}
\leq 4 \lambda\,\beta\: e^{\beta^2} \:.
\end{gather}
The resulting inequality can be estimated with the help of 
Lambert's $W$-function. Indeed, taking the square of the above inequality,
\[ \frac{k^2}{8 \lambda^2} \leq 2 \beta^2\: e^{2 \beta^2} \:, \]
one obtains (for details see~\cite[eq.~4.13.1]{dlmf})
\[ %\label{Imy0rel}
\beta^2 \geq \frac{1}{2}\: W\Big(\frac{k^2}{8 \lambda^2} \Big) \:. \]
In the region~$k \geq k_0$ under consideration, the argument of the $W$-function
is larger than~$e^2/2 \approx 3.69$, making it possible to use the inequalities
\[ \log x - \log \big( \log x \big) \leq W(x) \leq \log x \qquad \text{if~$x\geq \frac{e^2}{2}$}\:. \]
We thus obtain the estimate
\beq \label{imy0B2}
2\,\beta^2 \geq \log \Big(\frac{k^2}{8 \lambda^2} \Big) - \log \bigg( \log \Big(\frac{k^2}{8 \lambda^2} \Big) \bigg) \:.
\eeq
\end{itemize}
The different cases are shown schematically in Figure~\ref{figcases}.

We now state the main result of this section.
For notational convenience, 
\[ %\label{lesssimdef}
A \lesssim B \qquad \text{stands for} \qquad A \leq c\: B \]
for a suitable numerical constant $c>0$ (which does not depend on any parameters).

\begin{Prp} \label{prpges}
The function~$g(a,b)$ in~\eqref{gint} is bounded by
\begin{align}
|g(a,b)| &\lesssim e^{3a} \: e^{-h(\sqrt{2b},\lambda)} \label{gabes} \\
&= e^{3a} \:
\exp\bigg( \frac{3}{2} \beta^2 + \sqrt{2b} \:\Big( \frac{1}{2\beta} - \beta \Big) \bigg) \:,
\end{align}
where~$h$ is the function~\eqref{hdef}, and~$\beta$ is determined implicitly by~\eqref{krep} for~$k=\sqrt{2b}$, i.e.\
\beq \label{yhat}
\sqrt{2b} = 3\, \beta + 2 \lambda\,\beta\: e^{\beta^2}
\eeq
(and~$\lambda$ is given in terms of~$a$ by~\eqref{lambdadef}). More explicitly, $\beta$
is bounded from below by
\beq \label{imy0cases}
\beta \geq \left\{ \begin{array}{cl}
\displaystyle \frac{\sqrt{2b}}{3 + 2 e \lambda} & \text{in case~{\bf{(A)}}} \\[0.8em]
\displaystyle \frac{\sqrt{2b}}{6} & \text{in case~{\bf{(B1)}}} \\[0.8em]
\displaystyle \frac{1}{\sqrt{2}} \:\sqrt{\log \Big(\frac{b}{4 \lambda^2} \Big) - \log \bigg( \log \Big(\frac{b}{4 \lambda^2} \Big)}
 & \text{in case~{\bf{(B2)}}} \:,
\end{array} \right.
\eeq
with the cases as above with~$k=\sqrt{2b}$ and~$\beta$ given by~\eqref{yhat}.
\end{Prp}

We now enter the detailed estimates. The proof of this proposition will be completed at the end of this section.
Our strategy is to estimate the $k$-integral in the different regions separately. To this end, we decompose
the range of integration as
\[ (\sqrt{2b}, \infty) = I_{(A)} \dot{\cup} I_{(B1)} \dot{\cup} I_{(B2)} \]
with
\[ I_{A} = \big(\sqrt{2b}, k_0 \big)\:,\quad
I_{B1} = \big[\max \{\sqrt{2b}, k_0\}, k_1\big) \:,\quad
I_{B2} = \big[\max \{\sqrt{2b}, k_0, k_1\}, \infty \big) \:. \]
We begin with an estimate in case~{\bf{(A)}}.
\begin{Lemma} \label{lemmaA} The following inequality holds,
\[ g_{A} := \int_{I_{A}} \frac{k}{\sqrt{k^2-2b}}\:  \big|\hat{g}(a, k)\big| \: dk
\leq e^{3a} \:\exp\bigg( \sqrt{2b} \:\Big( \frac{1}{2\beta} - \beta \Big) \bigg) \:, \]
where~$\beta$ is chosen according to~\eqref{yhat}.
\end{Lemma}
\Proof Using the inequality~$0 \leq \beta<1$, we estimate~\eqref{ghes} by
\[ \big| \hat{g}(a,k) \big| \lesssim \frac{e^{3a}}{\sqrt{1 + \lambda}}\:
\exp \bigg( \lambda \, e^{\beta^2} \,\Big( 1 - 2\, \beta^2 \Big) \bigg) \:. \]
Setting~$x = \beta^2$, the last exponent involves the function
\beq \label{fdef2}
f(x):= e^x\, (1-2x) \:,
\eeq
whose first and second derivatives are negative,
\[ f'(x) = -e^x \,\big( 1+2x \big) < 0 \qquad \text{and} \qquad
f''(x) = -e^x \,\big( 3+2x \big) < 0 \:. \]
In particular, the function~$f$ is concave. Therefore, choosing~$\tilde{x}$, for all~$x>\tilde{x}$,
\[ f(x) \leq f\big( \tilde{x} \big) + f'(\tilde{x}) \, (x-\tilde{x}) \:. \]
As a consequence,
\[ \big| \hat{g}(a,k) \big| \lesssim \frac{e^{3a}}{\sqrt{1 + \lambda}}\:
\exp \bigg( \lambda \, f\big( \tilde{\beta}^2 \big) +
\lambda\, f'\big( \tilde{\beta}^2 \big) \: \big( \beta^2 - \tilde{\beta}^2 \big) \bigg) \:, \]
where we choose~$\tilde{\beta}$ such that~\eqref{yhat} holds.
Applying~\eqref{dkdimy0} and~\eqref{imy0A}, we obtain the estimate
\begin{align*}
\beta^2 - \tilde{\beta}^2 &= \int_{\tilde{k}}^k \frac{d}{dk'} \beta^2\: dk' 
= \int_{\tilde{k}}^k \frac{2 \beta}{3 + 2 \lambda\: e^{\beta^2}
+ 4 \lambda\: \beta^2\: e^{\beta^2}} \: dk' \\
&\geq \frac{2}{(3 + 2 e \lambda)(3 + 6 e \lambda)} \int_{\tilde{k}}^k k'\: dk' 
\geq \frac{1}{(3 + 6 e \lambda)^2} \: \big( k^2 - 2b \big) \:,
\end{align*}
where in the last line we also used that~$\beta<1$.
We thus obtain the estimate
\[ \big| \hat{g}(a,k) \big| \lesssim
\frac{e^{3a+\lambda f(\tilde{\beta}^2)}}{\sqrt{1 + \lambda}}\;
\exp \bigg( -\frac{\lambda\:|f'(\tilde{\beta}^2)|}{(3 + 2 e \lambda)^2} \: \big(k^2 - 2b \big) \bigg) \:. \]
Now we can estimate the integral by
\begin{align*}
g_A &\leq \int_{\sqrt{2b}}^{k_0} \frac{k}{\sqrt{k^2-2b}}\: \big| \hat{g}(a,k) \big| \:dk 
= \left\{ \begin{array}{c} z = \sqrt{k^2 - 2b} \\
z \,dz = k \, dk \end{array} \right\} \\
&= \int_0^{\sqrt{k_0^2-2b}} \big| \hat{g}(a,\sqrt{z^2+2b}) \big| \:dz \\
&\lesssim \frac{e^{3a+\lambda f(\tilde{\beta}^2)}}{\sqrt{1 + \lambda}}
\int_0^\infty \exp \bigg( -\frac{\lambda\:|f'(\tilde{\beta}^2)|}{(3 + 6 e \lambda)^2} \: z^2 \bigg)\:dz \\
&\lesssim
\frac{e^{3a+\lambda f(\tilde{\beta}^2)}}{\sqrt{1 + \lambda}}\: \frac{3+6 e \lambda}{\sqrt{\lambda\:|f'(\tilde{\beta}^2)|}}
\lesssim \frac{e^{3a+\lambda f(\tilde{\beta}^2_0)}}{|f'(\tilde{\beta}^2)|} \lesssim e^{3a+\lambda f(\tilde{\beta}^2)}\:,
\end{align*}
where in the last line we computed the Gaussian integral
and used that~$\lambda$ and~$|f'|$ are bounded from below.
Applying~\eqref{fdef2} and using that~$\tilde{\beta}<1$ gives the result
(where for notational convenience, in the statement of the lemma we omitted the tilde).
\QED

In order to estimate the integral in case~{\bf{(B)}}, we consider a general integral
\beq \label{gBdef}
g_B := \int_{\hat{k}}^{k_2} \frac{k}{\sqrt{k^2-2b}}\:  \big|\hat{g}(a, k)\big| \: dk
\eeq
with~$\hat{k} = \max \{ k_0, \sqrt{2b} \}$ and~$k_2 \geq \hat{k}$.
In this case, we write the estimate of Lemma~\ref{lemmages} using~\eqref{hes} as
\begin{align}
\big| \hat{g}(a,k) \big| &\lesssim \frac{e^{3a}}{\sqrt{1 + \lambda \, e^{\beta^2}}}\: e^{-h(\lambda,\hat{k})}
\: \exp \Big( -\hat{\beta}\: \big(k-\hat{k} \big) \Big) \notag \\
&\lesssim \frac{e^{3a}}{\sqrt{1 + \lambda \, e^{\tilde{\beta}^2}}}\: e^{-h(\lambda,\hat{k})}
\: \exp \Big( -\hat{\beta}\: \big(k-\hat{k} \big) \Big) \:, \label{ghB1B2}
\end{align}
(where in the last step we again used that~$\beta$ is  monotone increasing in~$k$).
In this inequality, the $k$-dependence is given simply by a decaying
exponential. Therefore, we may replace the upper limit of integration~$k_2$ in~\eqref{gBdef} by~$\infty$.
Thus it remains to estimate the integral
\[ \int_{\hat{k}}^\infty \frac{k}{\sqrt{k^2 - 2b}}\: e^{-\beta\: (k-\hat{k})} \: dk \:. \]
In preparation, we shift the integration variable such as to obtain an integral over the interval~$[\sqrt{2b}, \infty)$,
\begin{align}
&\int_{\hat{k}}^\infty \frac{k}{\sqrt{k^2 - 2b}}\: e^{-\beta\: (k-\hat{k})} \: dk = 
\left\{ k' = k - \ell \text{ with } \ell :=\hat{k} - \sqrt{2b} \geq 0 \right\} \notag \\
&=\int_{\sqrt{2b}}^\infty \frac{k' + \ell}{\sqrt{(k'+\ell)^2 - 2b}}\: e^{-\beta\: \big( k'-\sqrt{2b} \big)} \: dk' \notag \\
&\leq \int_{\sqrt{2b}}^\infty \frac{k'}{\sqrt{k'^2 - 2b}}\: e^{-\beta\: \big( k'-\sqrt{2b} \big)} \: dk' \:, \label{481a}
\end{align}
where in the last step we used that the integrand is monotone decreasing in~$\ell$.

\begin{Lemma} \label{lemmasqrtint}
For any parameters~$b \geq 0$ and~$d > 0$,
\begin{align*}
\int_{\sqrt{2b}}^\infty \frac{k}{\sqrt{k^2-2b}}\: e^{-d\, \big(k-\sqrt{2b} \big)}\: dk &\lesssim
\frac{b^\frac{1}{4}}{\sqrt{d}} + \frac{1}{d} \:.
%\\
%\int_{\sqrt{2b}}^\infty \frac{k}{\sqrt{k^2-2b}}\: k\: e^{-dk}\: dk &\lesssim
%\frac{b^\frac{3}{4}}{\sqrt{d}}\: e^{-\sqrt{2b}\, d}\:.
\end{align*}
\end{Lemma}
\Proof Introducing the variable~$z$ by
\[ z(k) := \sqrt{ \frac{k^2}{2b} - 1} \:,\qquad k = \sqrt{2b}\: \sqrt{z^2+1} \:,\qquad 
k\, dk = 2b\: z\, dz \:,\]
we obtain
\begin{align*}
\int_{\sqrt{2b}}^\infty \frac{k}{\sqrt{k^2-2b}}\: e^{-dk}\: dk
&= \int_0^\infty \frac{1}{\sqrt{2b}\: z}\: e^{-C \,\sqrt{z^2+1}}\:2b\, z\, dz\\
&= \sqrt{2b} \int_0^\infty e^{-C \,\sqrt{z^2+1}}\, dz
%\\
%\int_{\sqrt{2b}}^\infty \frac{k}{\sqrt{k^2-2b}}\: e^{-dk}\: k\: dk
%&= \int_0^\infty \frac{\sqrt{2b}\: \sqrt{z^2+1}}{\sqrt{2b}\: z}\: e^{-C \,\sqrt{z^2+1}}\:2b\, z\, dz\\
%&= 2b \int_0^\infty \sqrt{z^2+1}\: e^{-C \,\sqrt{z^2+1}}\, dz
\end{align*}
with
\[ C := d\: \sqrt{2b} \geq \sqrt{2} \:. \]
In order to estimate the integral further, we consider two cases:
\bitem
\item[(a)] $0 \leq z \leq 1$: The inequalities
\[ 1+\frac{z^2}{3} \leq \sqrt{z^2+1} \leq \sqrt{2} \]
give rise to the estimate
\begin{align*}
\int_{0}^1 e^{-C \,\sqrt{z^2+1}}\, dz
&\leq e^{-C} \int_{0}^1 e^{-\frac{C}{3}\: z^2}\: dz \\
&\leq e^{-C} \int_{0}^\infty e^{-\frac{C}{3}\: z^2}\: dz
= \frac{\sqrt{3 \pi}}{2}\: \frac{e^{-C}}{\sqrt{C}} \:.
%\\
%\int_{0}^1 \sqrt{z^2+1}\: e^{-C \,\sqrt{z^2+1}}\, dz
%&\leq \sqrt{2} \:e^{-C} \int_{0}^1 e^{-\frac{C}{3}\: z^2}\: dz
%%\\ &\leq \sqrt{2} \:e^{-C} \int_{0}^\infty e^{-\frac{C}{3}\: z^2}\: dz = 
%\leq \sqrt{\frac{3 \pi}{2}}\: \frac{e^{-C}}{\sqrt{C}} \:.
\end{align*}
\item[(b)] $1 \leq z$: In this case,
\[ \sqrt{2} + \frac{1}{\sqrt{2}}\: (z-1) \leq \sqrt{z^2+1} \leq \sqrt{2}\:z \:, \]
and thus
\begin{align*}
\int_1^\infty e^{-C \,\sqrt{z^2+1}}\, dz
&\leq e^{-C\, \sqrt{2}}\int_1^\infty e^{-\frac{C}{\sqrt{2}}\, (z-1)}\: dz
= e^{-C\, \sqrt{2}} \; \frac{\sqrt{2}}{C} \:.
%\\
%\int_1^\infty \sqrt{z^2+1}\: e^{-C \,\sqrt{z^2+1}}\, dz
%&\leq \sqrt{2} \:e^{-C\, \sqrt{2}}\int_1^\infty z\: e^{-\frac{C}{\sqrt{2}}\, (z-1)}\: dz 
%= 2 \,e^{-C\, \sqrt{2}} \Big( \frac{\sqrt{2}}{C^2} + \frac{1}{C} \Big) \:.
\end{align*}
\eitem
Collecting all the contributions gives the result.
\QED

\Proof[Proof of Proposition~\ref{prpges}]
Applying Lemma~\ref{lemmasqrtint} in~\eqref{gBdef}, \eqref{ghB1B2} and using~\eqref{481a}, we obtain
\[ |g_B| \lesssim \frac{e^{3a}}{\sqrt{1 + \lambda \, e^{\hat{\beta}^2}}}\: e^{-h(\lambda,\hat{k})}
\: \bigg( \frac{b^{\frac{1}{4}}}{\sqrt{\hat{\beta}}} + \frac{1}{\hat{\beta}} \bigg) \:.  \]
The terms in the denominator can be simplified because, using~\eqref{krep},
\[ \big( 1 + \lambda \, e^{\hat{\beta}^2} \big)\, \hat{\beta}
\simeq  \big( 3 + 2\lambda \, e^{\hat{\beta}^2} \big)\, \hat{\beta} = \hat{k} \:. \]
Applying~\eqref{halt}, we obtain the estimate
\beq \label{gBes}
|g_B| \leq e^{3a}\: \exp \bigg( \frac{3}{2}\: \hat{\beta}^2 + \frac{\hat{k}}{2\, \hat{\beta}} - \hat{\beta}\: \hat{k}
\bigg) \: \bigg( \frac{b^{\frac{1}{4}}}{\sqrt{\hat{\beta}}} + 1\bigg) \:,
\eeq
where we simplified the last summand inside the last brackets by using the inequality~$\hat{\beta} \geq 1$.
This concludes the estimates in case~{\bf{(B)}}.

Next, we need to add the integrals in the different regions.
Noting that~$\beta<1$ in case~{\bf{(A)}}, the estimate of Lemma~\ref{lemmaA} agrees
with the estimate in~\eqref{gBes} if we choose~$\hat{k}=\sqrt{2b}$.
Noting that, in view of~\eqref{dhdk}, the argument of the exponent is decreasing in~$\hat{k}$,
it suffices to consider the contribution in the region corresponding to the case determined by~$k=\sqrt{2b}$.
This gives~\eqref{gabes}.
The lower bounds in~\eqref{imy0cases} were derived in~\eqref{imy0A},
\eqref{imy0B1} and~\eqref{imy0B2}.
\QED

\subsection{Estimate of~\texorpdfstring{$g^{(2)}$}{gtwo}} \label{secsintegral}
We now come to the estimate of the solution of the Goursat problem~$g(a,b)$
in~\eqref{goursat} with initial data~$g_0^{(2)}$ as in~\eqref{g012}.
Our task is to estimate the~$s$-integral in~\eqref{g012}. In view of~\eqref{lambdadef},
this corresponds to integrating~$\lambda$ along a straight line
\[ \lambda = s^2\, \lambda_0 \qquad \text{with} \qquad s \in [0,1] \text{ and } \lambda_0 := e^{2a} \:. \]
More precisely, our task is to estimate the integral
\[ \int_0^1 |g(a,b)|\big|_{\lambda = s^2 \lambda_0}\: ds \]
with~$|g(a,b)|$ as estimated in~\eqref{gabes} and~$\beta$ as given implicitly by~\eqref{yhat}.

According to~\eqref{d2hdl2}, the function~$h(., \sqrt{2b})$ is convex. Hence
\[ h\big( \lambda, \sqrt{2b}) \geq h\big(\lambda_0, \sqrt{2b} \big)
+ \frac{\partial h\big(\lambda, \sqrt{2b}\big)}{\partial \lambda}\bigg|_{\lambda=\lambda_0} \: (\lambda-\lambda_0) \:. \]
As a consequence,
\begin{align*}
\int_0^1 & |g(a,b)|\Big|_{\lambda = s^2 \lambda_0}\: ds
\lesssim e^{3a} \int_0^1 e^{-h\big( s^2 \lambda_0, \sqrt{2b}\big)}\: ds \\
&\leq e^{3a} \int_0^1 e^{-h\big(\lambda_0, \sqrt{2b} \big) - \partial_\lambda  h\big(\lambda_0, \sqrt{2b} \big)\: \lambda_0\: (s^2-1)}\: ds \\
&= e^{3a}\: e^{-h\big(\lambda_0, \sqrt{2b} \big)}
\int_0^1 e^{-\partial_\lambda h\big(\lambda_0, \sqrt{2b} \big)\: \lambda_0\: (s^2-1)}\: ds \\
&=e^{3a}\: e^{-h\big(\lambda_0, \sqrt{2b} \big)}\: \frac{\sqrt{\pi}}{2} \frac{e^{-\nu}}{\sqrt{\nu}}\: \text{Erfi}(\nu) 
\end{align*}
with
\[ \nu := -\partial_\lambda h\big(\sqrt{2b}, \lambda_0 \big)\: \lambda_0 
\overset{\eqref{dhdl}}{=} \lambda \,e^{\beta^2}\big|_{\lambda=\lambda_0} \:, \]
where~$\text{Erfi}$ is the imaginary error function.

Using this result in the formula of Lemma~\ref{lemmaschwarz}, we obtain the following result:
\begin{Prp} \label{prpfinal} The solution of the Goursat problem~\eqref{goursat} with initial data~\eqref{g0def}
is bounded by
\begin{align*}
%\label{gabesfinal}
|g(a,b)| &\lesssim e^{3a} \:
\exp\bigg( \frac{3}{2} \beta^2 + \sqrt{2b} \:\Big( \frac{1}{2\beta} - \beta \Big) \bigg)
\: \sqrt{\frac{e^{-\nu}}{\sqrt{\nu}}\: \text{\rm{Erfi}}(\nu) } \:,
%&= e^{\frac{5a}{2}} \: \exp\bigg( \frac{5}{4} \beta^2 + \sqrt{2b} \:\Big( \frac{1}{2\beta} - \beta \Big) \bigg) \: \sqrt{e^{-\nu}\: \text{\rm{Erfi}}(\nu) } \:,
\end{align*}
where $\beta$ and~$\nu$ are given by
\begin{align*}
\sqrt{2b} &= 3\, \beta + 2 e^{2a}\,\beta\: e^{\beta^2} \\% \label{yhatfinal} \\
\nu &= e^{2a} \,e^{\beta^2} \:. %\label{nuhat}
\end{align*}
\end{Prp}

We finally state our results in a way compatible with Theorem~\ref{thmhaupt}.
\begin{Corollary} \label{cormain}
There is a numerical constant~$c>0$ such that the 
function~$R(\varepsilon, \omega)$ in~\eqref{main} can be chosen as
\beq \label{Rfinal}
R(\varepsilon, \omega) = c\:\exp\bigg( \frac{3}{2} \beta^2 + 2 \,\sqrt{|\log \varepsilon|}\:
\Big( \frac{1}{2\beta} - \beta \Big) \bigg) \: \sqrt{\frac{e^{-\nu}}{\sqrt{\nu}}\: \text{\rm{Erfi}}(\nu)}
\eeq
with~$\beta$ and~$\nu$ as given implicitly by
\begin{align}
2 \,\sqrt{|\log \varepsilon|} &= 3\, \beta + 8 \omega\,\beta\: e^{\beta^2} \label{yhatfinal2} \\
\nu &= 4 \omega \,e^{\beta^2} \:. \label{nuhat2}
\end{align}
\end{Corollary}
\Proof We use the result of Proposition~\ref{prpfinal} in Proposition~\ref{prop:series}
and apply~\eqref{abdef}.
\QED

We conclude this section with a brief discussion of our final result.
Clearly, due to the implicit definition of~$\beta$ and~$\nu$ via~\eqref{yhatfinal2} and~\eqref{nuhat2},
the estimate of Corollary~\ref{cormain} is rather involved. Its meaning can be revealed by
considering various limiting cases. For brevity, we here only consider a particular case which
explains why our last estimate goes beyond the previous estimates in 
Theorems~\ref{maintheoremsimple} and~\ref{thmmain2}. To this end, we consider the limiting case
\beq \label{logepsfinal}
|\log \varepsilon| \simeq \sqrt{\omega} \qquad \text{and} \qquad \omega \rightarrow \infty \:.
\eeq
In this limiting case, the first exponential inside the curly brackets in~\eqref{esprelim} 
is bounded from below, implying that the right side of~\eqref{esprelim} tends to infinity
as~$\omega \rightarrow \infty$. Thus Theorem~\ref{thmmain2} does not give any
information on the limiting case~\eqref{logepsfinal}.
On the other hand, the relation~\eqref{yhatfinal2} implies that~$\beta \sim \omega^{-3/4} \rightarrow 0$.
Consequently, \eqref{nuhat2} implies that~$\nu \sim \omega$, giving rise to an exponential
decay in~\eqref{Rfinal}. We conclude that Corollary~\ref{cormain} allows us to 
estimate~$R(\varepsilon, \omega)$ in
the limiting case~\eqref{logepsfinal}, although Theorem~\ref{thmmain2} fails.

\section{The \texorpdfstring{$3+1$}{three plus one}-Dimensional Case} \label{sec31dim}
Let~$B_1 \subset \R^3$ be the unit ball. We consider the Cauchy problem for the
scalar wave equation with smooth, compactly supported initial data in~$B_1$,
\[ \left\{ \begin{array}{c} (\partial_t^2 - \Delta_{\R^3}) \phi(t,\vec{x}) = 0 \\[0.3em]
\phi|_{t=0} = \phi_0 \in C^\infty_0(B_1) \:, \qquad \partial_t \phi|_{t=0} = \phi_1 \in C^\infty_0(B_1) \:.
\end{array} \right. \]
We denote the energy of the solution by
\beq\label{energy3d} E(\phi) := \frac{1}{2} \int_{B_1} \Big( 
\big|\partial_t \phi(0,\vec{x}) \big|^2 + \big| \nabla \phi(0,\vec{x})\big|^2 \Big)\: d^3x \:.
\eeq

In order to write the solution in an explicit form, it is useful to form the spatial Fourier transform
defined by
\[ \hat{\phi}(t,\vec{k}) = \int_{B_1} \phi(t,\vec{x}) \: e^{-i \vec{k} \vec{x}}\: d^3x \:. \]
Indeed, as is verified by direct computation, we have
\[ \hat{\phi}(t,\vec{k}) = \hat{\phi}_+(t,\vec{k}) + \hat{\phi}_-(t,\vec{k}) \]
with
\beq \label{phipmrep}
\hat{\phi}_\pm(t,\vec{k})
:= \frac{1}{2} \: e^{-i \omega t}\Big( \hat{\phi}_0(\vec{k}) \pm \frac{i}{\omega}\: \hat{\phi}_1(\vec{k}) \Big) \:,
\eeq
where we set
\[ %\label{omegadef2}
\omega = \omega(\vec{k}) := |\vec{k}| \:. \]
The solutions~$\phi_\pm$ are the components of positive and negative frequency, respectively.
We again express the energy with the help of Plancherel's theorem
as an integral in momentum space:
\begin{Lemma} \label{lemmaE331}
The energy~\eqref{energy3d} can be written as
\beq \label{epm4}
E(\phi) = E(\phi_+) + E(\phi_-) \qquad \text{with} \qquad
E_\pm(\phi) := \int_{\R^3} \frac{d^3k}{(2 \pi)^3} \:\omega^2 \: \big| \hat{\phi}_\pm(t, \vec{k}) \big|^2\:.
\eeq
\end{Lemma}
\Proof A direct computation using Plancherel's theorem gives
\begin{align*}
E(\phi) &= \frac{1}{2} \int_{\R^3} \frac{d^3k}{(2 \pi)^3} \:\Big( \omega^2\, \big|\hat{\phi}_0(\vec{k})\big|^2
+ \big|\hat{\phi}_1(\vec{k}) \big|^2 \Big) \\
&= \int_{\R^3} \frac{d^3k}{(2 \pi)^3} \;\omega^2\,\Big( \big|\hat{\phi}_+(t, \vec{k})\big|^2
+ \big|\hat{\phi}_-(t, \vec{k})\big|^2 \Big) \:,
\end{align*}
concluding the proof.
\QED

Due to spherical symmetry of the problem, we can expand the functions in spherical
harmonics, both in position and momentum space. For the initial data, we obtain
in polar coordinates~$(r, \vartheta, \varphi)$ the representations
\[ \phi_a(\vec{x}) = \sum_{l=0}^\infty \sum_{m=-l}^l Y_{lm}(\vartheta, \varphi)\: \phi_a^{lm}(r)
\qquad \text{with} \qquad a \in \{0,1\} \:. \]
Similarly, in momentum space we obtain the representations
\beq \label{phiaseries}
\hat{\phi}_a(\vec{k}) = \sum_{l=0}^\infty \sum_{m=-l}^l Y_{lm}(\vartheta, \varphi)\: \hat{\phi}_a^{lm}(\omega) \:,
\eeq
now in polar coordinates~$(\omega=|\vec{k}|, \vartheta, \varphi)$ in momentum space.
Since Fourier transformation preserves angular momentum, it follows that
the Fourier transformation of~$Y_{lm} \phi_a^{lm}$ is~$Y_{lm} \hat{\phi}_a^{lm}$.
Moreover, being the Fourier transform of functions supported in~$B_1(0)$, the
functions~$\hat{\phi}_a$ are real analytic. Therefore, they can be expanded in a
Taylor series about~$\vec{k}=0$. We write the resulting expansion as
\[ \hat{\phi}_a(\vec{k}) = \sum_{l=0}^\infty \sum_{m=-l}^l Y_{lm}(\vartheta, \varphi)
\sum_{p=0}^\infty c^{lm}_{a,p}\: \omega^{l+2p}\:. \]
In order to explain this formula, we note that the product~$Y_{lm}(\vartheta, \varphi)\:\, \omega^l$
is a homogeneous polynomial in~$\vec{k}$ of degree~$l$.
Therefore, in order to have a smooth function also in~$\omega$, the remaining series expansion
must involve only even powers of~$\omega$.
Using these expansions in~\eqref{phipmrep}, we obtain
\begin{gather}
\omega\, \hat{\phi}_\pm(t, \vec{k})
= e^{\mp i \omega t} \sum_{l=0}^\infty \sum_{m=-l}^l Y_{lm}(\vartheta, \varphi)\: \hat{h}^{lm}_\pm(\omega)
\quad \text{with} \label{phiexmom} \\
\hat{h}^{lm}_\pm(\omega) := \sum_{n=l}^\infty a_n^{lm}\: \omega^n \:, \label{hlmdef}
\end{gather}
where the coefficients are given by
\beq \label{almdef}
a^{lm}_{l+2p} = \pm \frac{i}{2}\, c^{lm}_{1,p} \qquad \text{and} \qquad
a^{lm}_{l+2p+1} = \frac{1}{2}\, c^{lm}_{0,p} \:.
\eeq
We point out that, in contrast to the $1+1$-dimensional case, here a parity splitting
is not necessary because it is already contained in the expansion in spherical harmonics
(indeed, even~$l$ corresponds to even parity, and odd~$l$ corresponds to odd parity).

In analogy to~~\eqref{sersplit}, the energies can be expressed in
terms of the functions~$\hat{h}^{lm}_\pm$ in~\eqref{hlmdef}:
\begin{Lemma} \label{lemmasplitangular}
The energies of the positive- and negative-frequency components of~$\phi$ in~\eqref{energy3d}
can be written as
\[ E(\phi_\pm) = \sum_{l=0}^\infty \sum_{m=-l}^l E^{lm}(\phi_\pm) \]
with
\beq \label{angularenergy}
E^{lm}(\phi_\pm)=E(Y_{lm}\, \phi_\pm^{lm})= \frac{1}{2 \pi^2}\int_0^\infty \bigg| 
\sum_{n=l}^\infty a^{lm}_n\, \omega^n \bigg|^2\: \omega^2\: d\omega \:
\eeq
\end{Lemma}
\Proof Using the expansion~\eqref{phiexmom}
in~\eqref{epm4} and using the orthonormality of the spherical harmonics, we obtain
\begin{align*}
E_\pm(\phi) &= \int_{\R^3} \frac{d^3k}{(2 \pi)^3} \:\omega^2 \: \big| \hat{\phi}_\pm(t, \vec{k}) \big|^2 \\
&= \sum_{l=0}^\infty \sum_{m=-l}^l \frac{4 \pi}{(2 \pi)^3}
\int_0^\infty \bigg| \sum_{n=l}^\infty a_n^{lm} \omega^n \bigg|^2 \:\omega^2\:d\omega \:.
\end{align*}
This concludes the proof.
\QED
We point out that there are two major differences compared to
the $1+1$-dimensional situation: First, the sum over~$n$ in~\eqref{hlmdef} starts at~$n=l$.
This is because the contributions of higher angular momentum vanish to higher order at~$k=0$.
Second and more importantly, the additional factor~$\omega^2$ in~\eqref{angularenergy}
is a result of the three-dimensional integration in polar coordinates in momentum space.

The next lemma gives an estimate of each Taylor coefficient in momentum space.
It can be regarded as the $3+1$-dimensional analog of Lemma~\ref{lemma1}.
\begin{Lemma}\label{lemmacoeff3d} Let~$\phi \in C^\infty_0(B_1)$ with angular decomposition
\[ \phi(x) = \sum_{l=0}^\infty \sum_{m=-l}^l Y_{lm}(\vartheta, \varphi)\: \phi^{lm}(r) \:. \]
Then its Fourier transform has a Taylor series representation
\[ \hat{\phi}(k) = \sum_{l=0}^\infty \sum_{m=-l}^l Y_{lm}(\vartheta, \varphi)
\sum_{p=0}^\infty c^{lm}_p\: \omega^{l+2p} \]
with coefficients bounded by
\begin{align}
|c^{lm}_p| &\leq \sqrt{\frac{4 \pi}{2l+1}}\: \frac{l!}{(2l-1)!!}\: \frac{1}{(l+2p)!}\:
\sqrt{\mu(B_1)}\: \|Y_{lm}\, \phi^{lm}\|_{L^2(B_1)} \label{cbound} \\
|c^{lm}_p| &\leq \sqrt{\frac{4 \pi}{2l+1}}\: \frac{l!}{(2l-1)!!}\: \frac{1}{(l+2p+1)!}\:
\sqrt{\mu(B_1)}\: \big\| \nabla \big( Y_{lm}\, \phi^{lm} \big) \big\|_{L^2(B_1)} \:. \label{cbound2}
\end{align}
\end{Lemma}
\Proof Since the Fourier transformation preserves angular momentum, it suffices to prove
the lemma for fixed~$l$ and~$m$. Moreover, by rotational symmetry
it suffices to consider the case~$m=0$
(more precisely, the transformation of the $m$-modes under rotations
is described by the Wigner $D$-matrix). Hence,
expressing the spherical harmonics in terms of Legendre polynomials
(see~\cite[eq.~14.30.1]{dlmf}), we obtain
\begin{align*}
\phi(x) &= Y_{l0}(\vartheta, \varphi)\: \phi^{l0}(r) \\
\hat{\phi}(k) &= Y_{l0}(\vartheta)
\sum_{p=0}^\infty c^{l0}_p\: \omega^{l+2p}
= \sqrt{\frac{2l+1}{4 \pi}}\; P_l(k_z) \sum_{p=0}^\infty c^{l0}_p\: |\vec{k}|^{2p}
\end{align*}
(where a factor~$\omega^l$ was absorbed into the Legendre polynomial).
In order to determine the coefficient~$c^{l0}_p$, we differentiate the last equation~$l+2p$ times
with respect to~$k_z$ and evaluate at~$k=0$,
\[ \big(\partial_{k_z}^{l+2p} \hat{\phi}\big)(0) = \begin{pmatrix} l+2p \\ l \end{pmatrix}\:
\sqrt{\frac{2l+1}{4 \pi}}\; P_l^{(l)}(0)\; c^{l0}_p\: (2p)! \:. \]
In order to compute the $l^\text{th}$ derivative of the Legendre polynomial, we must determine the
coefficient of its highest power. This can be accomplished with the help of the Rodrigues formula
(see~\cite[eq.~18.5.5]{dlmf})
\beq \label{Phighest}
P_l(x) = \frac{1}{2^l\, l!} \frac{d^l}{dx^l} \Big( (x^2-1)^l \Big)
= \frac{1}{2^l\, l!} \frac{d^l}{dx^l} \big( x^{2l} \big) + \O\big( x^{l-1} \big)
= \frac{1}{2^l\, l!} \: \frac{(2l)!}{l!} x^{l} + \O\big( x^{l-1} \big) \:,
\eeq
and differentiating~$l$ times gives
\[ P^{(l)}_l(0) = \frac{(2l)!}{2^l\, l!} = (2l-1)!!\:. \]
We thus obtain
\beq \label{e1}
\big(\partial_{k_z}^{l+2p} \hat{\phi}\big)(0)
= \sqrt{\frac{2l+1}{4 \pi}} \: (l+2p)!\: \frac{(2l-1)!!}{l!} \: c^{l0}_p \:.
\eeq
The partial derivative on the left can be estimated by
\begin{align*}
\big| \big(\partial_{k_z}^{l+2p} \hat{\phi}\big)(0) \big| &=
\bigg| \int_{B_1} (-i z)^{l+2p} \phi(\vec{x}) \: e^{-i \vec{k} \vec{x}}\: d^3x \bigg| \\
&\leq \int_{B_1} |\phi(\vec{x})|\: d^3x \leq \sqrt{\mu(B_1)}\: \|\phi\|_{L^2(B_1)} \:.
\end{align*}
Using this estimate in~\eqref{e1} and solving for~$c^{l0}_p$ gives~\eqref{cbound}.

In order to derive~\eqref{cbound2}, we again fix~$l$ and consider the case~$m=0$.
Differentiating~$\phi$ in the $z$-direction, we obtain
\begin{align*}
\widehat{\big( \partial_z \phi \big)}(k) &= k_z \, \hat{\phi}(k) = k_z\, Y_{l0}(\vartheta)
\sum_{p=0}^\infty c^{l0}_p\: \omega^{l+2p}
= \sqrt{\frac{2l+1}{4 \pi}}\; k_z \,P_l(k_z) \sum_{p=0}^\infty c^{l0}_p\: |\vec{k}|^{2p} \:.
\end{align*}
We now differentiate $l+2p+1$ times with respect to $k_z$ and evaluate at~$k=0$,
\[ \partial_{k_z}^{l+2p+1} \widehat{\big( \partial_z \phi \big)}(0)
= \begin{pmatrix} l+2p+1 \\ l+1 \end{pmatrix}\: \sqrt{\frac{2l+1}{4 \pi}}\; 
\partial_{k_z}^{l+1} \Big( k_z \,P_l(k_z) \Big) \Big|_{k_z=0}
\; c^{l0}_p\: (2p)! \:. \]
Again applying~\eqref{Phighest}, we obtain
\begin{align}
\partial_{k_z}^{l+2p+1} \widehat{\big( \partial_z \phi \big)}(0)
&= \begin{pmatrix} l+2p+1 \\ l+1 \end{pmatrix}\: \sqrt{\frac{2l+1}{4 \pi}}\; 
\frac{1}{2^l\, l!} \: \frac{(2l)!}{l!} \: (l+1)! \; c^{l0}_p\: (2p)! \notag \\
&= \sqrt{\frac{2l+1}{4 \pi}}\: (l+2p+1)! \: \frac{(2l-1)!!}{l!} \; c^{l0}_p \:. \label{511}
\end{align}
On the other hand, the partial derivative on the left can be estimated by
\begin{align*}
\big| \big(\partial_{k_z}^{l+2p+1} \widehat{\big( \partial_z \phi \big)}(0) \big| &=
\bigg| \int_{B_1} (-i z)^{l+2p+1} \,\big(\partial_z \phi\big)(\vec{x}) \: e^{-i \vec{k} \vec{x}}\: d^3x \bigg| \\
&\leq \int_{B_1} |\nabla \phi(\vec{x})|\: d^3x \leq \sqrt{\mu(B_1)}\: \|\nabla \phi\|_{L^2(B_1)} \:.
\end{align*}
Combining this estimate with~\eqref{511} gives~\eqref{cbound2}.
\QED

Similar to Proposition~\ref{prp1},
this lemma allows us to estimate each coefficient of the power series in~\eqref{hlmdef}.

\begin{Prp} \label{prp13}
The coefficients in the power series~\eqref{hlmdef} are bounded by
\beq \label{dldef}
|a^{lm}_n| \leq d_l\: \frac{\sqrt{E^{lm}(\phi)}}{n!} \qquad \text{with} \qquad
d_l := \frac{4 \pi}{\sqrt{6\,(2l+1)}}\: \frac{l!}{(2l-1)!!}\:.
\eeq
\end{Prp} 
\Proof Follows immediately by applying Lemma~\ref{lemmacoeff3d}
to the series~\eqref{phiaseries} and using~\eqref{almdef}.
More precisely, treating the cases of even and odd~$n$ separately, we obtain
\begin{align*}
\big| a^{lm}_{l+2p} \big| &= \frac{1}{2}\, \big|c^{lm}_{1,p}\big| 
\overset{\eqref{cbound}}{\leq} \frac{1}{2}\: \frac{\sqrt{6}}{\sqrt{4 \pi}}
\: \frac{d_l}{(l+2p)!}\: \sqrt{\mu(B_1)}\: \|Y_{lm}\, \phi_1^{lm}\|_{L^2(B_1)} \\
&= \frac{d_l}{(l+2p)!}\: \frac{1}{\sqrt{2}}\: \|Y_{lm}\, \phi_1^{lm}\|_{L^2(B_1)} 
\leq \frac{d_l}{(l+2p)!}\: \sqrt{E^{lm}(\phi)} \\
\big| a^{lm}_{l+2p+1} \big| &= \frac{1}{2}\, \big| c^{lm}_{0,p} \big| 
\overset{\eqref{cbound2}}{\leq} \frac{1}{2}\: \frac{\sqrt{6}}{\sqrt{4 \pi}}
\: \frac{d_l}{(l+2p+1)!}\:
\sqrt{\mu(B_1)}\: \big\| \nabla \big( Y_{lm}\, \phi_0^{lm} \big) \big\|_{L^2(B_1)} \\
&\leq \frac{d_l}{(l+2p+1)!}\: \frac{1}{\sqrt{2}}\: \big\| \nabla \big( Y_{lm}\, \phi_0^{lm} \big) \big\|_{L^2(B_1)}
\leq \frac{d_l}{(l+2p)!}\: \sqrt{E^{lm}(\phi)} \:.
\end{align*}
This concludes the proof.
\QED

We now use the same strategy as in Sections~\ref{sechighcoeff} and~\ref{secsmalltaylor}.
We decompose the series~$\hat{h}^{lm}_\pm$ in~\eqref{hlmdef} into a polynomial of degree~$N$
and the remainder term,
\beq \label{decompose3d}
\hat{h}^{lm}_\pm = \hat{h}^{lm}_N + R^{lm}_N 
\eeq
with
\[ \hat{h}^{lm}_N(\omega) := \sum_{n=l}^N a^{lm}_n \, \omega^n \qquad \text{and} \qquad
R^{lm}_N(\omega) := \sum_{n=N+1}^\infty a^{lm}_n \, \omega^n \:. \]
Similar to Lemma~\ref{lemmaremainder}, we first show that
the remainder term has small $L^2$-norm on the interval~$[0, \omega_1]$.
The main difference compared to Lemma~\ref{lemmaremainder} is the
additional factor~$\omega^2$ in the integration measure.
\begin{Lemma} \label{lemmaremainder3d}
Given~$\varepsilon \in [0, 1]$ and~$N \in \N_0$, we choose
\beq \label{omegamax2}
\omega_1 = \bigg( \frac{\varepsilon^2}{d_l^2}\: (N+1)!^2\, (2N+5) \bigg)^{\frac{1}{2N+5}} \:.
\eeq
Then the remainder term in~\eqref{decompose3d} is bounded on~$[0, \omega_1]$ by
\[ \|R^{lm}_{\pm\,N}(\omega)\|_{L^2([0, \omega_1], \,\omega^2 d\omega)} \leq  4\varepsilon\:\sqrt{E^{lm}(\phi)} \:. \]
\end{Lemma}
\Proof Applying Proposition~\ref{prp13}, we can estimate the remainder similar to~\eqref{cdef} by
\begin{align}
|R^{lm}_N(\omega)| &\leq d_l \sum_{n=N+1}^\infty \frac{\omega^n}{n!} \: \sqrt{E^{lm}(\phi^\eo)} \notag \\
&\leq d_l\, c(\omega)\: \frac{\omega^{N+1}}{(N+1)!} \:\sqrt{E^{lm}(\phi)}\qquad \text{with} \qquad c(\omega) := \sum_{n=0}^\infty \Big( \frac{\omega}{N+2} \Big)^n \:.
\label{cdef3}
\end{align}
Choosing~$\omega_1$ according to~\eqref{omegamax}, we know that
for $\varepsilon<1$ for all~$\omega \in [0,\omega_1]$,
\[ \frac{\omega}{N+2} \leq \frac{\omega_1}{N+2} \leq
\frac{\big( (N+1)!^2\, (2N+5) \big)^{\frac{1}{2N+5}}}{N+2} \leq \frac{3}{4}\:, \]
where the last inequality is verified by direct inspection and using the Stirling formula.
Therefore, the geometric series in~\eqref{cdef3} converges and is bounded by four,
\[ |R^{lm}_N(\omega)| \leq 4 d_l\: \frac{\omega^{N+1}}{(N+1)!} \:\sqrt{E^{lm}(\phi)} \:. \]
Using this pointwise bound, the $L^2$-norm can be estimated by
\begin{align*} \|R^{lm}_{\pm\,N}(\omega)\|^2_{L^2([0, \omega_1],\, \omega^2 d\omega)} &\leq 16 \,d_l^2\: E^{lm}(\phi)
\int_0^{\omega_1} \frac{\omega^{2N+4}}{(N+1)!^2}\: d\omega\\
&\leq
\frac{16 \,d_l^2\: E^{lm}(\phi)}{(N+1)!^2\, (2N+5)} \:\omega_1^{2N+5} \:,
\end{align*}
giving the result.
\QED

Now we can estimate each Taylor coefficient by using the method in Lemma~\ref{lemmapoly}.
The following result is the analog of Proposition~\ref{prpanes}.
\begin{Prp} \label{prpclmn}
Assume that for any given~$l \in \N_0$, $m \in \{-l, \ldots, l\}$ and~$\varepsilon \in (0,1]$,
\[ E^{lm}(\phi_-) \leq \varepsilon^2\: E^{lm}(\phi) \:. \]
Then the series coefficients in~\eqref{hlmdef} are bounded by
\[ |a^{lm}_n| \leq 25\: \max \Big(d_l, d_l^{\frac{2l+3}{2l+5}} \Big) \:\frac{1}{\sqrt{2n+1}} \:\frac{4^n}{n!}\: \varepsilon^{\frac{2}{2n+5}}
\: \sqrt{E^{lm}(\phi)}\:. \]
\end{Prp}
\Proof Given~$N \in \N_0$, we choose~$\omega_1$ as in~\eqref{omegamax2}. 
Decomposing the function~$\hat{h}^{lm}_-$ according to~\eqref{decompose3d},
the $L^2$-norm of the remainder is bounded according to Lemma~\ref{lemmaremainder3d}.
Combining this fact with Lemma~\ref{lemmasplitangular}, we obtain
\begin{align*}
&\|\hat{h}^{lm}_{N}(\omega)\|_{L^2([0, \omega_1],\, \omega^2 d\omega)}
= \big\| \hat{h}_-^{lm} - R_{N}^{lm} \big\|_{L^2([0, \omega_1],\, \omega^2 d\omega)} \\
&\leq \big\| \hat{h}_-^{lm} \big\|_{L^2([0, \omega_1],\, \omega^2 d\omega)}
+ \big\| R_{-\,N}^{lm} \big\|_{L^2([0, \omega_1],\, \omega^2 d\omega)}
\leq \sqrt{2\pi^2\, E^{lm}(\phi_-)} + \|R_{-\,N}^{lm}\|_{L^2([0, \omega_1])} \\
&\leq \varepsilon\, \sqrt{2\pi^2\, E^{lm}(\phi)} + 4\varepsilon\:\sqrt{E^{lm}(\phi)}
\leq 9 \varepsilon \, \sqrt{E^{lm}(\phi)} \:.
\end{align*}
Applying Lemma~\ref{lemmapoly} to the polynomial~${\mathcal{P}}(\omega):= \omega \,\hat{h}^{lm}_N(\omega)$ gives the bound
\begin{align*}
|a^{lm}_N|
&\leq \frac{1}{\sqrt{\omega_1}}\: \bigg( \frac{4}{\omega_1} \bigg)^{N+1}\:
\|{\mathcal{P}}\|_{L^2([0,\omega_1], d\omega)} 
= \frac{1}{\sqrt{\omega_1}}\: \bigg( \frac{4}{\omega_1} \bigg)^{N+1}\:
\|\hat{h}^{lm}_{N}(\omega)\|_{L^2([0, \omega_1],\, \omega^2 d\omega)} \\
&\leq 4^{N+1}\: \omega_1^{-N-\frac{3}{2}} \,6 \varepsilon \, \sqrt{E^{lm}(\phi)} \\
&\leq 9\cdot4^{N+1}\: d_l^{\frac{2N+3}{2N+5}} \:\varepsilon^{\frac{2}{2N+5}} \: (N+1)!^{-\frac{2N+3}{2N+5}}\: (2N+5)^{-\frac{2N+3}{4N+10}}\:
\sqrt{E^{lm}(\phi)} \:.
\end{align*}
The result follows asymptotically from the Stirling formula
and for small values of~$n$ directly by numerical evaluation.
\QED

Now we are ready to extend Proposition~\ref{prop:series} to the $3+1$-dimensional setting.
\begin{Prp} \label{prop:series4} Assume that for any given~$l \in \N_0$, $m \in \{-l, \ldots, l\}$ and~$\varepsilon \in (0,1]$,
the energy of the negative-frequency component is bounded in terms of the total energy by 
\[ E^{lm}(\phi_-) \leq \varepsilon^2\: E^{lm}(\phi) \:. \]
Then the initial data in momentum space is bounded pointwise for all~$\omega \in \R^+$ by
\[ %\label{initestimate}
\big|\hat{h}^{lm}(\omega)\big| \leq 25\: \max \Big(d_l, d_l^{\frac{2l+3}{2l+5}} \Big)\: \sqrt{E^{lm}(\phi)} \;
\big( 4\omega \big)^{-\frac{3}{2}}\: g_l\big(\omega, \varepsilon \big) \:, \]
where~$g_l$ is the series
\[ %\label{gseriesl}
g_l(\omega, \varepsilon) := \sum_{n=l}^\infty 
\frac{1}{\sqrt{2n+1}}\: \frac{(4 \omega)^{n+\frac{3}{2}}}{n!}\: \varepsilon^{\frac{2}{2n+5}} \:. \]
\end{Prp}

The series~$g_l$ in~\eqref{gseries} differ from the corresponding series~$g$ in~\eqref{gseries}
in two points: the sum begins at~$n=l$ (which makes the series smaller), and
the power of~$\varepsilon$ is $2/(2n+5)$ instead of~$2/(2n+3)$ (which makes the series larger).
The different power comes about as a consequence of the factor~$\omega^2$ in the integration
measure in~\eqref{angularenergy}.

The remaining task is to estimate the series~$g_l$. All the methods developed in
the $1+1$-dimensional setting can be adapted to the new series in~\eqref{gseries}.
A simple method for getting the connection is to estimate~$g_l$ by
\begin{align}
g_l(\omega, \varepsilon) &= \sum_{n=l}^\infty 
\frac{1}{\sqrt{2n+1}}\: \frac{(4 \omega)^{n+\frac{3}{2}}}{n!}\: \big( \varepsilon^{\frac{2n+3}{2n+5}} \big)^{\frac{2}{2n+3}}
\notag \\
%&\leq \sum_{n=l}^\infty  \frac{1}{\sqrt{2n+1}}\: \frac{(4 \omega)^{n+\frac{3}{2}}}{n!}\: \big( \varepsilon^{\frac{2l+3}{2l+5}} \big)^{\frac{2}{2n+3}} \notag \\
&\leq \sum_{n=0}^\infty 
\frac{1}{\sqrt{2n+1}}\: \frac{(4 \omega)^{n+\frac{3}{2}}}{n!}\: \big( \varepsilon^{\frac{2l+3}{2l+5}} \big)^{\frac{2}{2n+3}} 
= g\big( \omega, \varepsilon^{\frac{2l+3}{2l+5}} \big) \:. \label{gessimple}
\end{align}
This method is not quite optimal but seems sufficient for most applications.
For more refined estimates, one needs to reconsider the constructions in Sections~\ref{secgoursat}--\ref{secsintegral}
with modified exponents. For brevity, we do not enter the details here.

We conclude this section with two theorems. We begin with an estimate for each angular momentum
mode, obtained by combining Proposition~\ref{prop:series4} with the estimate~\eqref{gessimple} and Proposition~\ref{prpfinal}.

\begin{Thm} \label{thmmain3}
Let~$\phi(t,x)$ be a solution of the $3+1$-dimensional scalar wave equation which at some time~$t_0$ is
supported inside a ball of radius~$r>0$,
\[ \supp \phi(t_0, .) \in B_r(0) \:. \]
Assume that for any given~$l \in \N_0$, $m \in \{-l, \ldots, l\}$ and~$\varepsilon \in (0,1]$,
the energy of the negative-frequency component is bounded in terms of the total energy by 
\[ E^{lm}(\phi) \leq \varepsilon^2\: E^{lm}(\phi) \:. \]
Then there is an a-priori estimate for the momentum
distribution of~$\phi$ of the form
\[ %\label{main3}
\big|k\,\hat{\phi}^{lm}(k) \big| + \big| \partial_t \hat{\phi}^{lm}(k) \big| \leq R_l\big(\varepsilon, r \,|k| \big) \,\sqrt{r^3\,E^{lm}(\phi)}\:, \]
where the function~$R_l$ is given by
\[ R_l(\varepsilon, \omega) = c\: \max \Big(d_l, d_l^{\frac{2l+3}{2l+5}} \Big)\: 
\exp\bigg( \frac{3}{2} \beta^2 + \sqrt{2b} \:\Big( \frac{1}{2\beta} - \beta \Big) \bigg)
\: \sqrt{\frac{e^{-\nu}}{\sqrt{\nu}}\: \text{\rm{Erfi}}(\nu) } \:. \]
Here~$c$ is a numerical constant (which is independent of~$l$), $d_l$ are the constants in~\eqref{dldef},
and~$\beta$ and~$\nu$ are given implicitly by
\begin{align*}
2 \:\sqrt{\frac{2l+3}{2l+5}\: |\log \varepsilon|} &= 3\, \beta + 8 \omega\,\beta\: e^{\beta^2} \\% \label{yhatfinal3} \\
\nu &= 4 \omega \,e^{\beta^2} \:. %\label{nuhat3}
\end{align*}
\end{Thm}

Finally, by combining the estimates for each angular mode and summing over the modes,
we derive an estimate for a general solution to the $3+1$-dimensional wave equa\-tion.
\begin{Thm} \label{prop:series3} Assume that for~$\varepsilon \in (0,1]$,
the energy of the negative-frequency component is bounded in terms of the total energy by 
\[ E(\phi_-) \leq \varepsilon^2\: E(\phi) \:. \]
Then the $L^2$-norm of the spatial Fourier transform on a sphere of radius~$\omega$
is bounded for all~$\omega \in \R^+$ by
\[ %\label{initestimate}
\int_{S^2} \big|\omega\,\hat{\phi}(\vartheta,\phi,\omega)\big|^2\: d\mu_S^2(\vartheta, \varphi) \leq 625
\,d_0^{\frac{10}{3}}\, C\: E(\phi) \;
\big( 4\omega \big)^{-\frac{6}{2}}\: g^2_0\big(\omega, \varepsilon \big) \:, \]
where~$C$ is the constant
\[ %\label{Cseries}
C := \sum_{l=0}^\infty (2l+1) \,d_l^{\frac{4l+6}{2l+5}}<\infty \]
(and the~$d_l$ are again given by~\eqref{dldef}).
\end{Thm}
\Proof In order to simplify the calculations, we observe that $d_l>1$ only for $l=\{0,1,2,3\}$ and thus
\[ \max \Big(d_l, d_l^{\frac{2l+3}{2l+5}} \Big) \leq d_0^{\frac{5}{3}} \:d_l^{\frac{2l+3}{2l+5}} \qquad
\text{for all~$l\in \N_0$}\:. \]
Using this estimate in the statement of Proposition~\ref{prop:series4},
where we choose parameters~$\varepsilon_{lm}$ such that~$E^{lm}(\phi_-) = \varepsilon_{lm}^2\: E^{lm}(\phi)$,
we obtain
\begin{align*}
\int_{S^2} &\big|\omega \,\hat{\phi}(\vartheta,\varphi,\omega)\big|^2 \:d\mu_{S^2}
=\sum_{l=0}^\infty \sum_{m=-l}^l \big|\hat{h}^{lm}(\omega)\big|^2\\
&\leq  625\, d_0^{\frac{10}{3}}\: \sum_{l=0}^\infty \sum_{m=-l}^l  d_l^{\frac{4l+6}{2l+5}} \: E^{lm}(\phi)\big( 4\omega \big)^{-\frac{6}{2}}\: g^2_l\big(\omega, \varepsilon_{lm} \big) \:.
\end{align*}
Along the lines of the proof of Theorem \ref{fullsolution}, we use that the relations
\[ E^{lm}(\phi)= \delta_{lm}\: E(\phi) \qquad \text{and} \qquad E^{lm}(\phi_-)= \varepsilon^2_{lm}\: E^{lm}(\phi) \]
imply that for all $l,m$ with $\varepsilon_{lm}>\varepsilon$, the inequality~$\delta_{lm}\leq \frac{\varepsilon^2}{\varepsilon_{lm}^2}$ holds. We thus obtain
\begin{align*}
&\int_{S^2} \big|\omega\,\hat{\phi}(\vartheta,\varphi,\omega)\big|^2 \:d\mu_{S^2} \leq 625\, d_0^{\frac{10}{3}}\, E(\phi)\: \sum_{l=0}^\infty \sum_{m=-l}^l  d_l^{\frac{4l+6}{2l+5}} \:\delta_{lm}\big( 4\omega \big)^{-\frac{6}{2}}\: g^2_l\big(\omega, \varepsilon_{lm} \big)\\
    &\leq625\, d_0^{\frac{10}{3}}\, E(\phi)\: \left(\sum_{\varepsilon_{lm}\leq\varepsilon}d_l^{\frac{4l+6}{2l+5}} \:\big( 4\omega \big)^{-\frac{6}{2}}\: g^2_l\big(\omega, \varepsilon \big)  + \sum_{\varepsilon_{lm}>\varepsilon}d_l^{\frac{4l+6}{2l+5}} \:\big( 4\omega \big)^{-\frac{6}{2}}\: g^2_l\big(\omega, \varepsilon_{lm} \big)\frac{\varepsilon^2}{\varepsilon_{lm}^2}\right). 
\end{align*}
For all the modes with $\varepsilon_{lm}\leq\varepsilon$, we used that in this case, $g_l(\omega,\varepsilon_{lm})<g_l(\omega,\varepsilon) $ for all $l,m$, and that $\delta_{lm}\leq1$ due to Lemma~\ref{lemmasplitangular}.
With the same argument as in the proof of Theorem~\ref{fullsolution}, it follows that $\frac{\partial}{\partial \varepsilon_{lm}} \bigg(
g_l^2(\omega,\varepsilon_{lm}) \:\frac{\varepsilon^2}{\varepsilon_{lm}^2} \bigg)<0 $ for $\varepsilon \in [0,1)$ and thus 
\[ g_l^2(\omega,\varepsilon_{lm}) \:\frac{\varepsilon^2}{\varepsilon_{lm}^2}\leq g_l^2(\omega,\varepsilon)
\qquad \text{for all~$l,m$}\:, \]
giving rise to the estimate
\begin{align*}
    \int_{S^2}   \big|\omega\,\hat{\phi}(\vartheta,\varphi,\omega)\big|^2 \:d\mu_{S^2} & \leq    625\, d_0^{\frac{10}{3}}\, E(\phi)\: \sum_{l=0}^\infty \sum_{m=-l}^l d_l^{\frac{4l+6}{2l+5}} \:\big( 4\omega \big)^{-\frac{6}{2}}\: g^2_l\big(\omega, \varepsilon \big)\\
    & \leq    625\, d_0^{\frac{10}{3}}\, E(\phi)\big( 4\omega \big)^{-\frac{6}{2}}\: g^2_0\big(\omega, \varepsilon \big)\: \sum_{l=0}^\infty \sum_{m=-l}^ld_l^{\frac{4l+6}{2l+5}} \:,
\end{align*}
where in the last step we used that $g_l(\omega,\varepsilon)\leq g_0(\omega,\varepsilon)$ for all $l\in \N$.
Carrying out the sum over $m$, we obtain the series
\[ \sum_{l=0}^\infty (2l+1) \: d_l^{\frac{4l+6}{2l+5}} \:. \]
Using~\eqref{dldef} and applying Stirling's formula to each term of the resulting series
\[ \sum_{l=0}^\infty (2l+1)^{\frac{2}{2l+5}} \:\bigg(\frac{8 \pi^2}{3}\: \Big( \frac{l!}{(2l-1)!!}\Big)^2
\bigg)^{\frac{2l+3}{2l+5}} \:, \]
one sees that this series converges absolutely. This completes the proof. 
\QED

\appendix
\section{Alternative Derivation of the Integral Representation} \label{appA}
In this appendix, we give an alternative derivation of the integral representation
of the solutions of the Goursat problem~\eqref{gabform}.
The method is by direct computation using the series representation of the Bessel function~$J_0$.

\begin{Lemma} \label{lemmaconvolution} Let~$g(a)$ be a power series of the form
\[ g(a) = \sum_{n=0}^\infty c_n\: e^{(2n+3)a} \:. \]
Then for all~$a,b>0$,
\beq \label{convolution}
\sum_{n=0}^\infty c_n\: e^{(2n+3)a - \frac{b}{2n+3}} = \int_{-\infty}^a J_0 \Big( 2 \,\sqrt{(a-\tau)\, b}\, \Big)\:
g'(\tau)\: d\tau\:.
\eeq
\end{Lemma}
\Proof The Bessel function~$J_0$ has the power expansion (see~\cite[eq.~10.2.2]{dlmf})
\[ J_0(z) = \sum_{\ell=0}^\infty \frac{(-1)^\ell}{(\ell!)^2}\: \Big( \frac{z^2}{4} \Big)^\ell \:. \]
Denoting the right side of~\eqref{convolution} by~$T(a,b)$, we obtain
\begin{align*}
T(a,b) \,&\!:= \int_{-\infty}^a J_0 \Big( 2 \,\sqrt{(a-\tau)\, b} \,\Big)\: g'(\tau)\: d\tau \\
&= \int_{-\infty}^a \sum_{\ell,n=0}^\infty \frac{(-1)^\ell}{(\ell!)^2}\: \big( (a-\tau)\, b \big)^\ell\;
(2n+3)\: c_n\: e^{(2n+3)\,\tau} \: d\tau \:.
\end{align*}
Introducing the new integration variable~$\xi=(2n+3)(a-\tau)$ gives
\begin{align*}
T(a,b) &= \int_0^\infty \sum_{\ell,n=0}^\infty \frac{(-1)^\ell}{(\ell!)^2}\: \Big( \frac{b}{2n+3}\: \xi \Big)^\ell\;
(2n+3)\: c_n\: e^{-\xi + (2n+3)\,a} \: \frac{d\xi}{2n+3} \\
&= \sum_{\ell,n=0}^\infty \frac{(-1)^\ell}{(\ell!)^2}\: \Big( \frac{b}{2n+3} \Big)^\ell\;
c_n\: e^{(2n+3)\,a} \: \int_0^\infty  \xi^\ell\, e^{-\xi} \,d\xi \\
&= \sum_{\ell,n=0}^\infty \frac{(-1)^\ell}{\ell!}\: \Big( \frac{b}{2n+3} \Big)^\ell\; c_n\: e^{(2n+3)\,a} 
= \sum_{n=0}^\infty \exp \Big(-\frac{b}{2n+3} \Big)\; c_n\: e^{(2n+3)\,a} \:,
\end{align*}
where in the last step we carried out the $\ell$-series to obtain an exponential.
\QED

\Thanks {{\em{Acknowledgments:}}
C.F.P.\ was supported by the Australian Research Council grant DP170100630.
Further, a part of the work has been supported by the Swedish Research Council under grant no.~2016-06596 while
the author was in residence at Institut Mittag-Leffler in Djursholm, Sweden, during the fall semester of 2019. C.F.P.\ was also funded by the SNSF grant P2SKP2 178198. We are grateful for support by the Vielberth Foundation, Regensburg. Finally, we would like to thank the referees for their constructive feedback.

%\bibliographystyle{amsplain}
%\bibliography{../felix}
\providecommand{\bysame}{\leavevmode\hbox to3em{\hrulefill}\thinspace}
\providecommand{\MR}{\relax\ifhmode\unskip\space\fi MR }
\providecommand{\MRhref}[2]{%
  \href{http://www.ams.org/mathscinet-getitem?mr=#1}{#2}
}
\providecommand{\href}[2]{#2}

\end{document}